\documentclass[aps,prd,twocolumn,superscriptaddress,nofootinbib,longbibliography]{revtex4-2}

\usepackage[english]{babel}
\usepackage{graphicx}
\usepackage{dcolumn}
\usepackage{bm}
\usepackage{amsmath,amssymb,amsfonts,amsthm}
\usepackage{dsfont}
\usepackage{float}
\usepackage{array}
\makeatletter
\let\newfloat\newfloat@ltx
\makeatother
\usepackage{algpseudocode}
\usepackage{multirow}
\usepackage{colortbl}
\usepackage{booktabs}
\usepackage{stmaryrd}
\usepackage{verbatim}
\usepackage[normalem]{ulem}
\usepackage[utf8]{inputenc}
\usepackage{siunitx}
\usepackage{listings}
\usepackage{xcolor}
\usepackage{empheq}
\usepackage{braket}
\usepackage[colorlinks = true,
            linkcolor = blue,
            urlcolor  = blue,
            citecolor = blue,
            anchorcolor = blue]{hyperref}
\usepackage{comment}
\usepackage{orcidlink}
\usepackage{tcolorbox}
\usepackage{fontawesome5}
\usepackage{makecell}

\newcolumntype{C}{>{$}c<{$}}
\newcolumntype{P}[1]{>{\centering\arraybackslash}p{#1}}
\newcolumntype{M}[1]{>{\centering\arraybackslash}m{#1}}

\AtBeginDocument{%
  \heavyrulewidth=.08em
  \lightrulewidth=.05em
  \cmidrulewidth=.03em
  \belowrulesep=.65ex
  \belowbottomsep=0pt
  \aboverulesep=.4ex
  \abovetopsep=0pt
  \cmidrulesep=\doublerulesep
  \cmidrulekern=.5em
  \defaultaddspace=.5em
}

\newcommand{\be}{\begin{equation}}
\newcommand{\ee}{\end{equation}}

\newcommand{\bA}{\boldsymbol{A}}
\newcommand{\bM}{\boldsymbol{M}}
\newcommand{\bK}{\boldsymbol{K}}
\newcommand{\bSigma}{\boldsymbol{\Sigma}}
\newcommand{\bQ}{\boldsymbol{Q}}
\newcommand{\bW}{\boldsymbol{W}}
\newcommand{\bGamma}{\boldsymbol{\Gamma}}
\newcommand{\bn}{\boldsymbol{n}}
\newcommand{\bd}{\boldsymbol{d}}

\newcommand{\bx}{\boldsymbol{x}}
\newcommand{\bw}{\boldsymbol{w}}
\newcommand{\bB}{\boldsymbol{B}}
\newcommand{\bP}{\boldsymbol{P}}
\newcommand{\btheta}{\boldsymbol{\theta}}
\newcommand{\blambda}{\boldsymbol{\lambda}}

\newcommand{\bI}{\mathds{1}}

\newcommand{\tr}{\operatorname{Tr}}
\newcommand{\Stm}{S_{\rm tm}}
\newcommand{\Soms}{S_{\rm oms}}
\newcommand{\mHz}{\,{\rm mHz}}
\newcommand{\Hz}{\,{\rm Hz}}

\newcommand{\lisanoiserepo}{https://github.com/uk-lisa-gs/lisa-noise}
\newcommand{\lisanoisetag}{main}

\newcommand{\paperdoi}{10.5281/zenodo.23504465}
\newcommand{\lisanoisedoi}{10.5281/zenodo.23501674}
\DeclareRobustCommand{\srclink}[1]{%
  \href{https://doi.org/\paperdoi}{\faCode\,\texttt{\detokenize{#1}}}}

\begin{document}

\title{Noisy Pipelines I: An end-to-end pipeline to perform multivariate noise inference for LISA}

\author{Ollie~Burke\,\orcidlink{0000-0003-2393-209X}}\email{ollie.burke@glasgow.ac.uk}
\affiliation{School of Physics and Astronomy, University of Glasgow, Glasgow G12 8QQ, UK}

\author{Joseph~Bayley\,\orcidlink{0000-0003-2306-4106}}
\affiliation{School of Physics and Astronomy, University of Glasgow, Glasgow G12 8QQ, UK}

\author{Charlie~Hoy\,\orcidlink{0000-0002-8843-6719}}
\affiliation{Institute of Cosmology and Gravitation, University of Portsmouth,
Portsmouth, PO1 3FX, UK }

\author{Laura~K.~Nuttall\,\orcidlink{0000-0001-7472-0201}}
\affiliation{Institute of Cosmology and Gravitation, University of Portsmouth,
Portsmouth, PO1 3FX, UK }
\begin{abstract}

Instrumental data from the Laser Interferometer Space Antenna (LISA) must be post-processed prior to any global-fit based analysis. For example, it is critical to filter, downsample, segment and window the data in order to perform inference of both the signal and noise parameters in the frequency domain. Each step alters the noise process, requiring a correction to the noise covariance matrix that describes the underlying data stream. We present \texttt{lisa-noise}, a JAX-based noise inference pipeline that returns a spectral estimate of the noise process subject to post-processing effects. We fit splines to the underlying secondary noises and fit the full multivariate $3 \times 3$ covariance matrix of the Michelson channels. We assume unequal and non-constant arm-lengths throughout. The drift of the zeros of the transfer functions, a non-stationary effect, is captured by averaging the transfer functions over a number of epochs within each segment. On 30 days of conditioned simulated data, a likelihood evaluation costs $3\,$ms on a current laptop for a 7 day segment. A gradient-based fit from a tight prior recovers the injected spectra in a fraction of a second. A multi-start search over large priors $\pm 10$ in $\log_{10}$ centred on the simulated truth obtains a noise estimate consistent with the data, verified by Bayesian techniques. The fitted covariance whitens the conditioned data in every band and through every zero of the transfer functions, in a form that global-fit pipelines can use directly.

\end{abstract}

\date{\today}
\maketitle

\section{Introduction}\label{sec:intro}

The Laser Interferometer Space Antenna (LISA) is a future large class mission pioneered by the European Space Agency (ESA)~\cite{LISA2017,LISA:2024hlh}, with the goal of probing the mHz frequency band of sources with mission duration spanning four to ten years. 
Data analysis problems for LISA
have significant challenges: many overlapping resolvable (and unresolvable) GW sources~\cite{Cornish:2005qw,umstatter2005lisa,Crowder_2007,Robson:2017ayy,Littenberg:2020bxy,Karnesis:2021tsh,Littenberg:2023xpl}, a moving constellation with evolving arm-lengths~\cite{Cornish:2003tz,Shaddock:2003dj,Tinto:2003vj,tinto2014time,Martens:2021phh,Hartwig:2021mzw,Bayle_2023} and data likely plagued with gaps~\cite{Carre:2010ra,Baghi:2019eqo,Dey:2021dem,Blelly:2021oim,Burke:2025bun,Castelli:2024sdb,burke_tdi_gaps,burke2026mindsecondgapimpact} and noise transients~\cite{Robson:2018jly,Baghi:2021tfd,LISAPathfinder:2022awx,Spadaro:2023muy,Houba:2024tyn,Muratore:2025knh,sauter2025maximumlikelihooddetectioninstrumental,Boumerdassi:2025gvf}. These latter three points contribute to the overall non-stationary element of the noise process, inducing time-dependence in the statistical elements describing the noise. Under simplified assumptions, community-driven global-fit pipelines have demonstrated success with (accurately) extracting a plethora of entangled GW sources in the presence of Gaussian and stationary noise~\cite{Littenberg:2020bxy, Katz:2024oqg, GeeMoo}, with inference placed on the signals and the noise simultaneously, as a space-based detector with no signal-free data requires.

The global-fit pipelines~\cite{Littenberg:2020bxy, Katz:2024oqg, GeeMoo} were tested on the \emph{Sangria} data set~\cite{LDC_Sangria_zenodo}, whose noise is a random draw from a known covariance matrix, Gaussian, stationary and circulant. \textsc{Erebor}~\cite{Katz:2024oqg} and \textsc{Gee-Moo}~\cite{GeeMoo} assume the spectral shape and fit the two amplitudes of the two fundamental noises, the Optical Metrology System (OMS) and the Test Mass acceleration (TM) noise. \textsc{GLASS}~\cite{Littenberg:2023xpl} adapts the trans-dimensional \textsc{BayesLine} algorithm~\cite{Littenberg:2014oda} with an Akima spline~\cite{akima1970new}, each knot carrying a frequency and a level. Strub \emph{et al.}~\cite{Strub:2024kbe} fit a spline to a smoothed Welch estimate. None of these had to account for the spectral artefacts of transforming raw time-domain data to the frequency domain, and with equal, constant arms their noise channels were orthogonal. Outside the global fit, two lines of work estimate the LISA noise from time-domain simulations of the instrument. The first estimates the spectral matrix of the Time Delay Interferometry (TDI) channels directly, with no model of the instrument, through penalised B-splines for a single channel~\cite{Aimen:2025zxn} and, for the full $3\times 3$ matrix of $X$, $Y$, $Z$, P-splines on a Cholesky factor~\cite{Vajpeyi:2026pzh}, a matrix-gamma process prior~\cite{Liu:2026ydn} and a variational approximation that scales to a year of data~\cite{Liu:2026yzk}. These make no assumption about the noise sources, show that treating $A$, $E$, $T$ as independent fails once the optical benches differ~\cite{Vajpeyi:2026pzh}, and use constant light travel times. The second line, which we follow, models the single-link noises and propagates them through the known TDI transfer functions. Baghi \emph{et al.}~\cite{Baghi:2023qnq} fitted one spline to a common single-link spectrum, with one transfer function for all noises, on a year of simulated data with unequal, time-varying arms, for stochastic-background detection~\cite{Hartwig:2023pft,Kume:2024xvh,Santini:2025iuj}. 

To our knowledge, none of the works above carries the conditioning of the data, beyond the normalisation of the window function, into the noise covariance, or models the drift of the zeros of the TDI transfer functions within an analysed segment of data. A spline fitted directly to the TDI spectra absorbs the conditioning for free. It has no link to the instrument, which is the route of the companion paper~\cite{Hoy:2026aaa}. Here we fit the two single-link spectra through the known TDI kernels, so the conditioning must be modelled. The two approaches are complementary, and we expect the eventual solution to be a hybrid of them, with instrumental input where it exists and an agnostic spline for the rest. 

In this work we introduce two pipeline codes. The first, \texttt{MojitoProcessor}~\cite{Burke_MojitoProcessor_2026}, accepts instrumental noise generated with \texttt{lisainstrument}~\cite{bayle2023lisainstrument} and converted to TDI variables with \texttt{pytdi}~\cite{Staab2022PyTDI}, and conditions the data. It band-pass filters to the frequency band of interest $f \in [10^{-4},10^{-1}]\,$Hz, downsamples to reduce the data volume, cuts the record into segments and tapers them to mitigate spectral leakage. Every step changes the noise process. The second, \texttt{lisa-noise}~\cite{lisa_noise_2026}, receives the conditioned segments and estimates the two secondary noise spectra, with the conditioning and the known (evolving) light travel times (LTTs) of the instrument folded into the $3\times 3$ noise covariance matrix of $\{X,Y,Z\}$. We model the noise at the single-link level and propagate it through the known TDI transfer functions, as in Ref.~\cite{Baghi:2023qnq}. The arm-lengths of the constellation change over a segment. We account for this by invoking local stationarity and averaging the numerical TDI transfer functions over the light travel times within the segment. Only the spectral shapes of the two secondary noises are free, and these are the quantities we estimate. 

The remainder of the manuscript is organised as follows. Section~\ref{sec:data} propagates the single-link noises to the TDI spectra and states the likelihood. Section~\ref{subsec:simulation} describes the simulation and the conditioning. Section~\ref{sec:modelled} develops the noise model and the inference, from the splines through the averaged kernels and the covariance of the conditioned data to the likelihood and the inference schemes. Section~\ref{sec:results} collects the results and Sec.~\ref{sec:conclusions} concludes. The appendices give the derivations, supplementary verification material and the corner plots representing the sampled posterior.

\section{LISA data and the noise process}\label{sec:data}

\subsection{From single links to TDI}\label{subsec:links_to_tdi}

Time Delay Interferometry is an algebraic technique used to suppress the most dominant of LISA's noise sources: laser frequency noise~\cite{tinto2014time}. 
The LISA data stream of channels $\mathcal{I} \in \{X, Y, Z\}$ is modelled as
\begin{equation}\label{eq:data_model}
    d^{(\mathcal{I})}(t) = h^{(\mathcal{I})}(t;\btheta) + n^{(\mathcal{I})}(t;\blambda)\,,
\end{equation}
with $h^{(\mathcal{I})}$ a GW contribution governed by source parameters $\btheta$ on top of an instrumental noise process $n^{\mathcal{I}}$ governed by $\blambda$. Our priority throughout this manuscript is to estimate the noise process $n$, so we set $h = 0$ and leave any form of signal analysis to future work.

The TDI variables are constructed using the six intermediate measurements $\eta_{ij}$, each a specific combination of delayed interferometer measurements~\cite{Bayle:2019dfu}. In our analysis, we assume that the six optical metrology system (OMS) noises and six test-mass noises are independent, zero-mean, Gaussian and wide-sense stationary. Our analysis assumes that there are only two secondary noise sources remaining in LISA data after TDI has been performed. 
The noise spectra of both are governed by their Power Spectral Densities $S_{\rm tm}$ and $S_{\rm oms}$. From~\cite{Hartwig:2021mzw}, we can write the covariance of single-link measurements as
\begin{subequations}\label{eq:link_cov}
\begin{align}
    \big[\bSigma_{\eta}\big]_{ij,\,ij} &= \frac{2\Delta t}{N}\,\mathbb{E}\big[\tilde{\eta}_{ij}\tilde{\eta}_{ij}^{*}\big] = 2\,\Stm^{\rm ff} + \Soms^{\rm ff}\,,\\
    \big[\bSigma_{\eta}\big]_{ij,\,ji} &= \frac{2\Delta t}{N}\,\mathbb{E}\big[\tilde{\eta}_{ij}\tilde{\eta}_{ji}^{*}\big] = \Stm^{\rm ff}\,\varphi_{ij}(f)\,,\\
    \varphi_{ij}(f) &= e^{\,2\pi i f d_{ji}} + e^{-2\pi i f d_{ij}}\,.
\end{align}
\end{subequations}
Here $d_{ij}$ denotes the light travel time (LTT) along the arms, with light travelling from spacecraft $j$ to $i$,
$\Delta t$ is the cadence and $N$ the number of samples. Here we write the two spectra of interest $\Stm$ in $\mathrm{m^2\,s^{-4}\,Hz^{-1}}$ and $\Soms$ in $\mathrm{m^2\,Hz^{-1}}$, which enter the $\eta$ variables in fractional frequency units (ff)
\begin{equation}\label{eq:unit_conversion}
    \Stm^{\rm ff}(f) = \frac{\Stm(f)}{(2\pi f c)^2}\,, \qquad \Soms^{\rm ff}(f) = \Soms(f)\,\Big(\frac{2\pi f}{c}\Big)^{2}\,,
\end{equation}
with parametric forms 
\begin{align}
    \Stm^{\rm seed}(f) &= A_{\rm tm}^{2}\left[1 + \Big(\frac{f_{\rm knee,1}}{f}\Big)^{2}\right]\left[1 + \Big(\frac{f}{f_{\rm knee,2}}\Big)^{4}\right], \label{eq:psd_seed_tm}\\
    \Soms^{\rm seed}(f) &= A_{\rm oms}^{2}\left[1 + \Big(\frac{f_{\rm knee,3}}{f}\Big)^{4}\right]\,.\label{eq:psd_seed_oms}
\end{align}
We will refer to these as the ``seed'' PSDs, where the noise spectra parameters feed into the instrument simulator discussed in Sec.~\ref{subsec:simulation} with simulation parameters in Table~\ref{tab:sim_params}. 

Following~\cite{Baghi:2023qnq}, we project the six single link variables $\boldsymbol{\eta}$ to three TDI variables using the mixing matrix $\bM$~\cite{Hartwig:2023pft}. The mixing matrix is defined in Appendix~\ref{app:mixing_matrix_inst_cov}, Eq.~(\ref{eq:mixing_explicit}--\ref{eq:mixing_N}). This mixing matrix accounts for the complex exponential factors that appear due to transforming delayed time-series into the frequency domain. It is possible to construct the linear map 
\begin{equation}\label{eq:mixing_def}
    \tilde{\bd}(f) = \bM(f;\{d_{ij}\})\,\tilde{\boldsymbol{\eta}}(f)\,, \qquad \bM \in \mathbb{C}^{3\times 6}\,.
\end{equation}
The mixing matrix depends on the frequency $f$ and on the six light travel times $d_{ij}$ only. We can write
\begin{equation}
    \mathbb{E}\big[\tilde{\bd}\tilde{\bd}^{\dagger}\big] = \frac{N}{2\Delta t}\,\bM\bSigma_{\eta}\bM^{\dagger} \equiv \frac{N}{2\Delta t}\,\bSigma_{\rm inst}\,,
\end{equation}
for $\bSigma_{\rm inst}$ the covariance of the noise process specific to the instrumental properties (derived through TDI computations). A calculation detailed in App.~\ref{app:mixing_matrix_inst_cov} reveals that the instrumental covariance matrix is made up of a product of the test-mass and readout power spectra and their respective TDI transfer functions
\begin{empheq}[box=\fbox]{equation}\label{eq:kernel_factorisation}
    \bSigma_{\rm inst}(f) = \Stm^{\rm ff}(f)\,\bK_{\rm tm}(f) + \Soms^{\rm ff}(f)\,\bK_{\rm oms}(f)\,,
\end{empheq}
with the two TDI transfer functions $\bK_{\rm tm}$ and $\bK_{\rm oms}$, functions of frequency and LTTs at some instant $d_{ij}(t)\approx d_{ij}(t_0)$,
\begin{subequations}\label{eq:kernels}
\begin{align}
    \bK_{\rm oms} &= \bM\bM^{\dagger}\,,\\
    \bK_{\rm tm} &= 2 \bM\bM^{\dagger} + \sum_{(ij,\,ji)} \Big(\varphi_{ij}\, \bM_{:,ij}\,\bM_{:,ji}^{\dagger} + {\rm h.c.}\Big)\,,
\end{align}
\end{subequations}
The kernels $\bK$ are $3\times 3$ Hermitian matrices at every frequency which encode the geometry of the constellation. The two TDI transfer functions carry no dependence on the noise parameters and are determined by the orbit of the LISA constellation. The delay factors of $\bM$ are oscillatory functions of both the light travel time between arms and frequency. From Eq.~\eqref{eq:mixing_explicit} the zero of channel $X$ is given at the frequency
\begin{equation}\label{eq:frequency_zeros}
f = \frac{k}{d_{12} + d_{21} + d_{13} + d_{31}} \approx \frac{k}{4\bar{d}}\,.
\end{equation}
where $\bar{d}$ is the mean one-way light travel time of the four links at spacecraft 1 at an instant of time $t$ and $k$ some positive integer. The light travel times therefore dictate the locations of the zeros as a function of frequency (see Fig.~\ref{fig:data_vs_instrument}). There are infinitely many zeros, but the zeros of interest for our purposes are the three that lie in the science band $f\in (10^{-4}, 10^{-1})\,$Hz, i.e., $k \in \{1,2,3\}$. For a mean LTT of 8.33 seconds, the zeros are located at $f = \{30, 60, 90\}\,$ mHz.

The evolution of the arm-lengths forces the locations of the zeros to move by a select number of frequency bins of bin width $\Delta f = 1/T$. For the $k^{\rm th}$ zero, the zeros move by
\begin{equation}\label{eq:drift_equation}
\frac{\delta f_k}{\Delta f} = -k \left(\frac{T}{4\bar{d}}\right)\left(\frac{\delta \bar{d}}{\bar{d}}\right) \sim T^2.
\end{equation}
Here $\delta \bar{d} = [\bar{d}(t_0 + T) - \bar{d}(t_0)] \sim T$ is how much the mean changes between the start and end of the time segment. The number of bins by which a zero drifts is related to (1) the length of the segment duration, (2) the variation in the arm-lengths (through $\delta \bar{d}$) and (3) the specific zero of interest $k \in \{1,2,3\}$. 

Since the TDI combinations are formed on the ground, the form of the mixing matrix $\bM$ is known in advance. The only quantity here not known in advance (with respect to the kernels) are the link LTTs. Our analysis assumes that the LTTs are known perfectly. In reality, there will be uncertainty in the LTTs, since the arm-lengths should be known at best to within $\sim 1\,$m~\cite{Bayle:2019dfu}. With $\bar{d} = 2.5\times 10^{9}/c$ and $\delta\bar{d} = 1/c$, and assuming a  $T = 7\,$ day segment, the number of bins the zeros drift due to uncertainty in the LTTs is $\delta f_1 / \Delta f \sim 10^{-5}$. The measured LTTs are not a limiting factor, only how much the arm-lengths change over a specific segment. We will return to this in Sec.~\ref{subsec:simulation}.

\subsection{Local stationarity and the likelihood}\label{subsec:likelihood}

Our analysis is conducted in the frequency domain where we approximate each segment as a locally stationary process. Instead of treating the Fourier domain data $\tilde{\boldsymbol{d}} = [\tilde{\boldsymbol{d}}^{X}, \tilde{\boldsymbol{d}}^{Y}, \tilde{\boldsymbol{d}}^{Z}]^T \in \mathbb{C}^{3\times N_f}$, we instead stack the vector by Fourier bins 
\begin{equation}\label{eq:stacked_data_vector}
\tilde{\boldsymbol{d}}_{j} = [d^{(X)}(f_j), d^{(Y)}(f_j), d^{(Z)}(f_j)] \in \mathbb{C}^{3}
\end{equation}
for $j \in \{0,1,2,\ldots, N_f - 1\}$. The Fourier components of a stationary Gaussian process are independent across frequency, so the covariance of the stacked data vector in \eqref{eq:stacked_data_vector} is block diagonal with $3\times 3$ blocks $\propto \bSigma(f_j)$. As Appendix~\ref{app:fast_inv_det} shows, re-ordering of the TDI data streams as above results in a covariance that is the direct sum $\bigoplus_j (N/2\Delta t)\,\bSigma(f_j)$. In other words, a block-diagonal matrix. This is essential for efficiency of our log-likelihood computations, allowing for vectorised matrix operations.  

For our analysis, we choose to use the Whittle-based likelihood~\cite{whittle:1957,Finn:1992wt}
\begin{equation}\label{eq:whittle}
    \ln\mathcal{L}(\blambda) = -\sum_{j \in \mathcal{S}} \Big[\ln\det\bSigma_j + \tfrac{2\Delta t}{N}\, \tilde{\bd}_j^{\dagger}\bSigma_j^{-1}\tilde{\bd}_j\Big] + {\rm const}\,,
\end{equation}
with $\bSigma_j(\blambda) \equiv \bSigma(f_j;\blambda) \in \mathbb{C}^{3 \times 3}$ the model covariance matrix the analyst decides to use evaluated at a single frequency bin $j$. Here $\mathcal{S}$ denotes the set of retained bins, which in our case are the bins with $f_j \in [10^{-4}, 10^{-1}]\Hz$. 

Equation~\eqref{eq:whittle} is only exact when the underlying noise covariance matrix is circulant. A circulant matrix is diagonalisable via the Discrete Fourier Transform matrix, yielding a purely diagonal frequency domain covariance matrix with entries proportional to the Power Spectral Density. Even if the properties of the process were circulant, then segmenting the data returns a Toeplitz matrix. As described in Appendix B of Ref.~\cite{Burke:2025bun}, this results in small off-diagonal frequency domain elements that we choose to ignore in this work. A Toeplitz matrix is asymptotically (in terms of number of samples) equivalent to a circulant one~\cite{gray2006toeplitz}. In other words, for long durations of data the covariance can be treated as circulant for parameter inference.  

Here Eq.~\eqref{eq:whittle} is written in the per-bin-component of $\{X,Y,Z\}$ with the full $3\times 3$ matrix, accounting for correlations between channels. Our inference scheme focuses on estimating $S_{\rm oms}$ and $S_{\rm tm}$ through computing these noise spectra and converting to a TDI noise covariance matrix, Eq.~\eqref{eq:kernel_factorisation}. The data then feed into Eq.~\eqref{eq:whittle} to compute the likelihood. The computational cost of Eq.~\eqref{eq:whittle} is given in the first row of Fig.~\ref{fig:cost}. 

\subsection{Simulated data and post-processing}\label{subsec:simulation}

We generate a noise-only data set with the instrument simulator \texttt{lisainstrument}~(v2.2)~\cite{bayle2023lisainstrument,Bayle_2023}. We will assume unequal and time-varying arms, where the individual spacecraft positions are propagated along the numerically built ESA trailing orbits built into the \texttt{lisaorbits}~(v3.0)~\cite{Bayle_lisaorbits} package. The output telemetry are converted to second generation TDI observables in $\{X,Y,Z\}$ via the \texttt{pytdi}~\cite{Staab2022PyTDI} package. These TDI variables are computed only in terms of fluctuations of beat notes, rather than total frequencies. The top panel of Table~\ref{tab:sim_params} collects the simulator settings.

For our simulation, we neglect tilt-to-length coupling noise, clock noise, laser modulation noise and sideband noise, and assume perfect cancellation of the laser frequency noise. We also neglect pseudo-ranging noise, so that the LTTs used throughout TDI are those within the \texttt{ESA-orbits} file. We refer the reader to Refs.~\cite{Hartig_2022, PhysRevApplied.14.014030,PhysRevD.103.123027} for more literature on these sources of noise which we neglect. Noise levels across OMS and TM acceleration are assumed equal across spacecraft. 

The TDI streams, the light travel times between arms and the spacecraft orbits are written in the level 1 (L1) file format using the \texttt{mojito} package~\cite{mojito_2026}. 

\begin{table}[t]
\centering
\caption{The data set. \emph{Top}: the simulated L1 data, the TDI streams as the instrument model delivers them. \emph{Bottom}: the post-processed L1 data, the product of \texttt{MojitoProcessor} (Sec.~\ref{subsec:simulation}) that every analysis in this paper consumes; each setting is described in the package documentation~\cite{Burke_MojitoProcessor_2026}. The L1 rate is $f_1 = \SI{4}{Hz}$; the decimated rate is $f_s = \SI{0.2}{Hz}$, whose Nyquist frequency $f_s/2 = \SI{0.1}{Hz}$ is the top of the analysis band.}
\label{tab:sim_params}
\footnotesize
\setlength{\tabcolsep}{4pt}
\begin{tabular}{@{}ll@{}}
\toprule
\multicolumn{2}{@{}l@{}}{\textbf{L1 data} (simulated instrument)} \\
\midrule
Duration              & \SI{30}{days} \\
L1 sampling      & $f_1 = \SI{4}{Hz}$ ($\Delta t_1 = \SI{0.25}{s}$), $N_1 = 10\,368\,000$ \\
Orbits                & ESA trailing (\texttt{lisaorbits} v3.0) \\
Light travel times    & 8.29--8.35\,s, unequal, evolving \\
Laser locking         & six-laser, static frequency plan \\
Test-mass noise       & $3\,\mathrm{fm\,s^{-2}\,Hz^{-1/2}}$ \\
                      & $f_{\rm knee,1} = 0.4\mHz$, $f_{\rm knee,2} = 8\mHz$ \\
Readout (OMS) noise   & $15\,\mathrm{pm\,Hz^{-1/2}}$; $f_{\rm knee,3} = 2\mHz$ \\
TDI variables         & Michelson $X_2,Y_2,Z_2$ (2nd generation) \\
TDI precision         & extended (80-bit \texttt{longdouble}) \\
\midrule
\multicolumn{2}{@{}l@{}}{\textbf{L1 post-processed} (\texttt{MojitoProcessor})} \\
\midrule
Units                 & fractional frequency, $\nu_0 = \SI{2.816e14}{Hz}$ \\
Band-pass filter      & Butterworth, order 2, zero-phase \\
                      & corners $50\,\mu$Hz and \SI{0.16}{Hz} \\
Decimation            & $4 \to \SI{0.2}{Hz}$ ($D = 20$) \\
                      & Kaiser $\beta = 31$ FIR, 401 taps \\
Trim                  & \SI{1}{day} at each end \\
Segments              & $4 \times \SI{7}{days}$; $N = 120\,960$, $\Delta t = \SI{5}{s}$ \\
Taper                 & Tukey, $\alpha = 0.05$, per segment \\
Analysis band         & $[10^{-4}, 10^{-1}]\Hz$, $N_f = 60\,420$ bins \\
                      & $\Delta f = \SI{1.65}{\micro Hz}$ \\
Product               & HDF5: $X,Y,Z$ segments, light travel \\
                      & times, a record of every setting above \\
\bottomrule
\end{tabular}
\end{table}

The \texttt{MojitoProcessor} package~\cite{Burke_MojitoProcessor_2026} reads in the three TDI data streams, orbital information and the sampling properties. It then gives the user the freedom to filter, downsample, trim, segment, and window the data stream for downstream analyses. The package returns the time-domain segmented post-processed data, the light travel times, and metadata describing the post-processing effects (filtering, decimation, windowing, segmentation). The details of the package are carefully outlined in the documentation of Ref.~\cite{Burke_MojitoProcessor_2026} and we refer the interested reader there. Our post-processing simulation parameters are outlined in the bottom block of Table~\ref{tab:sim_params}. 

\begin{figure}[t]
    \centering
    \includegraphics[width=\linewidth]{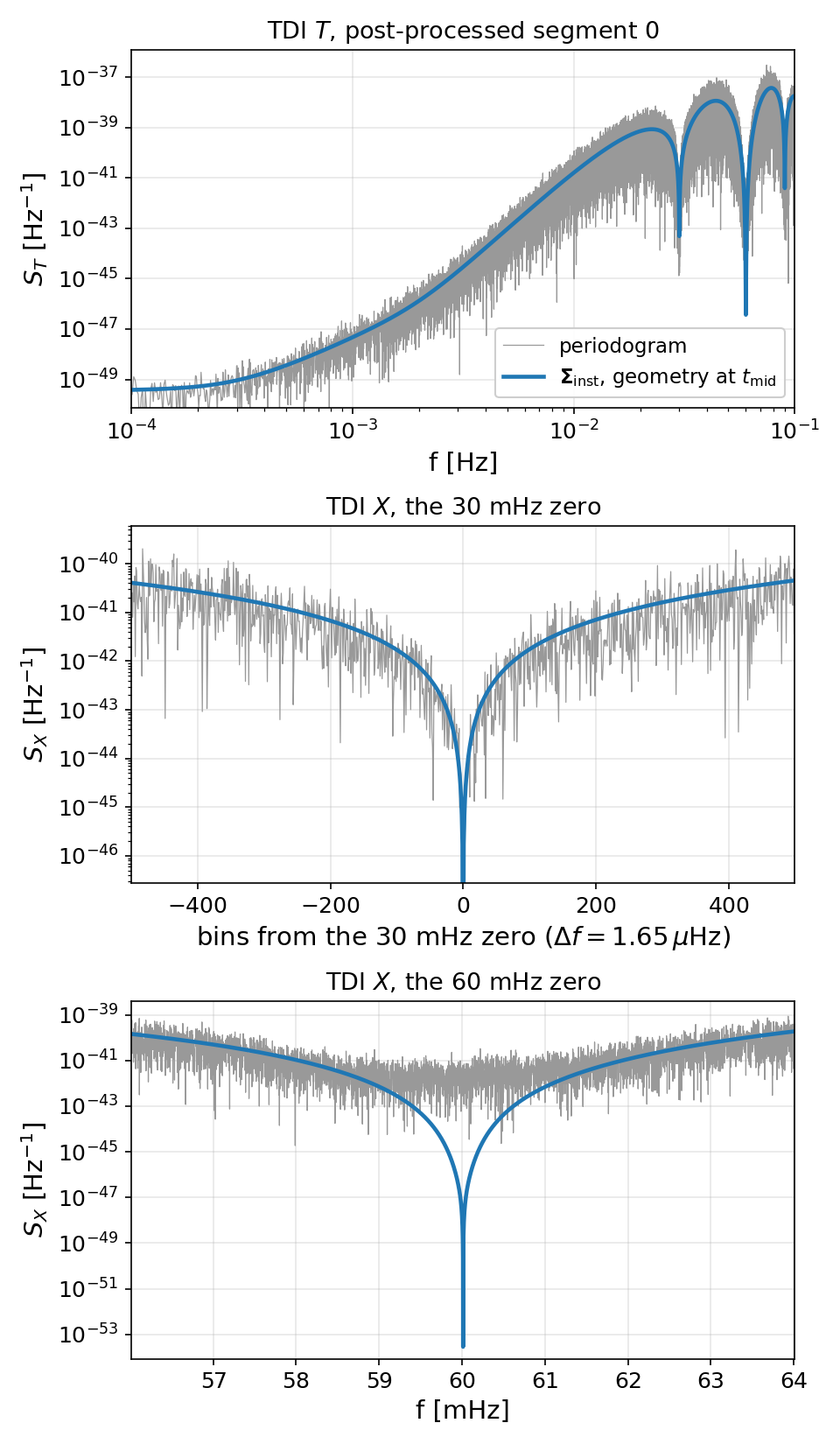}
    \caption{(Top panel): The post-processed data of the $T$ channel of the first segment against the instrument covariance $\bSigma_{\rm inst}$ of Eq.~\eqref{eq:kernel_factorisation} at the injected spectra, with LTTs evaluated at the midpoint of the segment $t = t_{\rm mid}$. (Middle panel): The first zero of $X$ at \SI{30}{mHz} with $500$ bins to either side. The zero of the data drifts by eight bins over the week and is smeared over them. The frozen covariance falls to zero at one bin. The bottom panel is $X$ at the \SI{60}{mHz} zero, \SI{4}{mHz} to either side. The decimation folds power from above the Nyquist frequency onto this zero. The instrument covariance is significantly inconsistent with the data at this zero. Produced by \srclink{fig_conditioning.py}.}
    \label{fig:data_vs_instrument}
\end{figure}

Figure~\ref{fig:data_vs_instrument} (top) shows the periodogram of $T$ on the first segment of data, segment 0, against the instrument PSD\footnote{An artefact of the filters, the taper or the decimation that is invisible in $X$, $Y$, $Z$ appears in $T$ as excess power at a few $\times 10^{-4}\Hz$, and we have used it throughout as a diagnostic of the conditioning. We remind the reader that inference is performed on $\{X,Y,Z\}$, but $T$ still stands as a useful diagnostic of the post-processing effects.} from \textsc{segwo} using Eq.~\eqref{eq:kernel_factorisation}. Since $T$ is a linear combination of the Michelson channels $\{X,Y,Z\}$, artefacts that arise in $T$ will also be present in each of $\{X,Y,Z\}$, hence why we show $T$ over the full band rather than a single Michelson channel. Here the instrumental covariance is evaluated at the simulation parameters of Table~\ref{tab:sim_params}, with the LTTs at $t = t_{\rm mid}$, the middle of the segment. By eye the two agree over the whole band, away from the zeros. The lower two panels of Fig.~\ref{fig:data_vs_instrument} give a closer look at the first two zeros of the $X$ channel. 

From the orbit file, over the first segment of data with $T_{\rm seg} = 7\,$ days the arms from spacecraft 1 shorten by $\delta \bar{d} = -3.7\,$ms, with $\bar{d} = 8.33\,$s. Substituting into Eq.~\eqref{eq:drift_equation} gives
\begin{equation}
\frac{\delta f_{k}}{\Delta f} \approx 8k\,\text{bins}
\end{equation}
implying that the first, second and third zeros drift by 8, 16 and 24 bins respectively over the segment duration. At \SI{30}{mHz}, the covariance Eq.~\eqref{eq:kernel_factorisation}, evaluated at $t = t_{\rm mid}$ falls to zero at a single bin and does not take into account the 8 smeared bins over the segment. At \SI{60}{mHz} (\SI{90}{mHz}), not only have the zeros drifted by $\sim 16 (24)$ bins, but the decimation procedure has folded power from above the Nyquist frequency onto this zero. None of these features are built into the instrument covariance. Section~\ref{sec:avg_TDI_response} averages the geometry over the segment. Section~\ref{subsec:window} folds the taper, the filters and the aliased power into the covariance. Without these post-processing and drifting bin corrections the estimated spectra would be biased and would propagate to the astrophysical parameters, yielding both imprecise and inaccurate parameter estimates. 

\section{Noise Inference Engine}\label{sec:modelled}\label{sec:missing_data_section}

In this section, we describe our noise inference scheme. First we will discuss our spline treatment on the secondary noises and second how we use the \emph{averaged} TDI responses to project the single-link measurements into TDI observables. We will then describe the impact of conditioning the data using \texttt{MojitoProcessor} and how the noise covariance matrix can be corrected for it. We discuss computational aspects of the likelihood and various inference schemes -- ranging from maximum likelihood estimation to full Bayesian inference. 

\subsection{Secondary noises and their spline representation}\label{subsec:splines}

We represent each single-link spectrum as a cubic spline in the log--log plane, as in Ref.~\cite{Baghi:2023qnq},
\begin{equation}\label{eq:spline}
    \log_{10} S_{\alpha}(f) = \sum_{i=1}^{n_{\alpha}} B^{\alpha}_{i}(\log_{10} f)\,\lambda^{\alpha}_{i} + b^{\alpha}_{\rm ghost}(f)\,,
\end{equation}
for $\alpha \in \{{\rm tm}, {\rm oms}\}$, where the free parameters $\lambda^{\alpha}_{i}$ are the $\log_{10}$ values of the spectrum at the knots, $B^{\alpha}_{i}(f)$ is the cardinal basis of the spline (the spline that is one at knot $i$ and zero at every other knot) and $b^{\alpha}_{\rm ghost}$ collects the contribution of two \emph{ghost} knots placed outside the band, by default at $0.7\,f_{\min}$ and $1.5 \, f_{\max}$, and held at their seed values so that they anchor the spline at the band edges; they are never inferred. 

\begin{figure}[t]
    \centering
    \includegraphics[width=\linewidth]{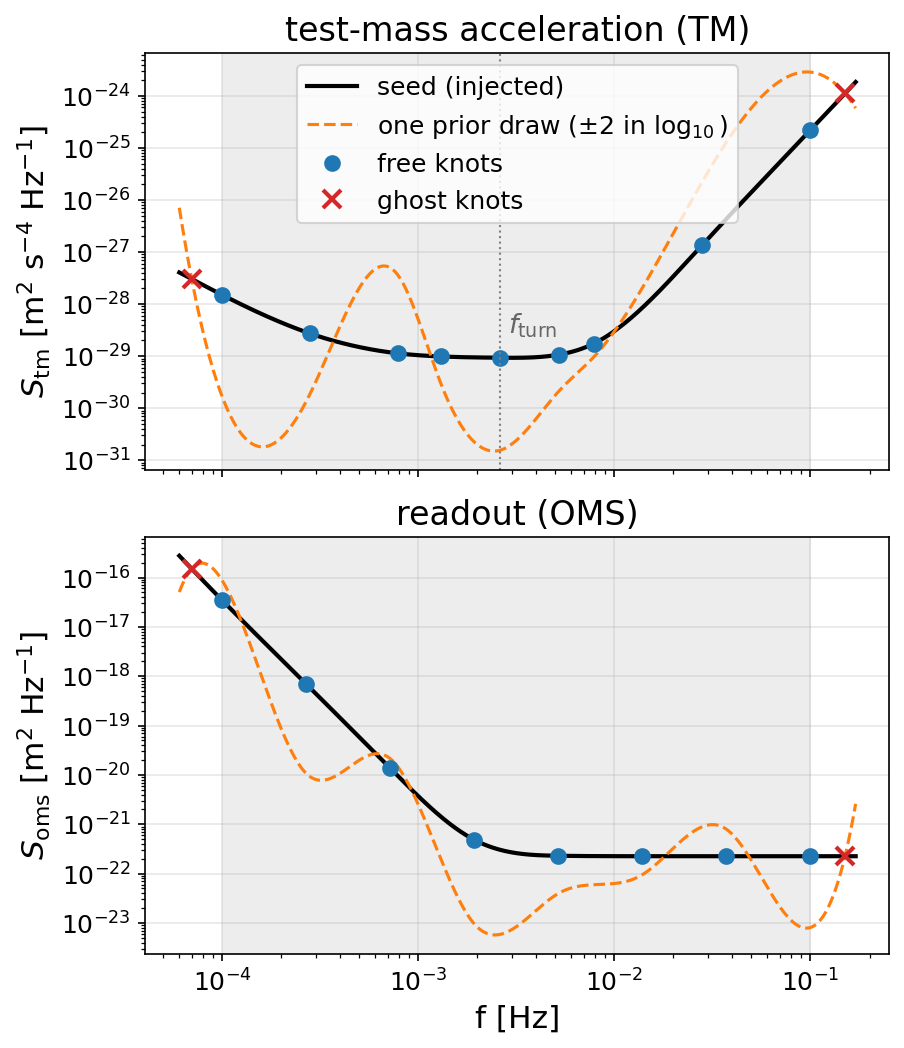}
    \caption{The spline representation of the two secondary noises. The seed spectra of Eqs.~\eqref{eq:psd_seed_tm}--\eqref{eq:psd_seed_oms}, in requirement units, acceleration for the test mass and displacement for the readout. The free knots (filled circles) and the fixed ghost knots outside the band (crosses). The analysis band is shaded. The test-mass knots cluster around the turning point of the spectrum at $f_{\rm turn} \simeq 2.6\mHz$ whereas the readout knots are uniform in $\log f$. The dashed curve is one draw from the uniform prior on the knot values, narrowed to $\pm 2$ in $\log_{10}$ for legibility. For inference we consider $\pm 10$ in $\log_{10}$ for the prior. Produced by \srclink{fig_mle_only.py}.}
    \label{fig:model_ingredients}
\end{figure}

Unlike~\cite{Baghi:2023qnq,Littenberg:2014oda,Littenberg:2023xpl}, we do not inform the knot placements from the data using inference techniques (such as reversible jump MCMC (RJMCMC)). Instead, for simplicity, we use prior knowledge of the instrument (through Eqs.~\eqref{eq:psd_seed_tm} and \eqref{eq:psd_seed_oms}) to place the knots. Prior to inference, the knots are placed once and are common to all segments. We discuss this knot placement below. 
For the OMS noise, whose fractional-frequency spectrum is a smooth power law, eight knots are spaced uniformly in $\log f$ across the band. For the test-mass noise, whose spectrum in displacement units has a minimum at $f_{\rm turn} \simeq 2.6\mHz$ with red/blue noise at low/high frequencies, we use a more informed knot placement. Nine knots are placed, three below the turning point spaced in $\log f$, three at $(0.5, 1, 2)\,f_{\rm turn}$ and three above. Through trial and error, with this specific spectral shape, we have found that this knot placement fits the data sufficiently well without over/under fitting the spectra. With the ghosts, that is ten and eleven knots respectively for OMS and test mass, with $n_{\lambda} = 9 + 8 = 17$ the number of free parameters per segment. An example of the knot placement is shown in Figure~\ref{fig:model_ingredients}. The knot count is a modelling choice rather than a fitted quantity, unlike in Refs.~\cite{Littenberg:2014oda,Littenberg:2023xpl}.  

For inference we regard the analytical model of Eqs.~\eqref{eq:psd_seed_tm}--\eqref{eq:psd_seed_oms} as the truth set by the simulation parameters in Table~\ref{tab:sim_params}. We construct a cubic spline on the truth, with the knot placements described in the above paragraph and compute the true knot values given the true simulation parameters. For inference, the prior on each knot value is set to be uniform over $\pm 10$ in $\log_{10}$ about the seed. With a uniform prior this wide, the posterior maximum coincides with the maximum of the likelihood. 
  
We remark that we are using knowledge of the truth to build the spline and set the ghost points. In essence, the true PSD only sets the centre of the prior box we sample over. We believe this is reasonable as experts related to LISA instrumentation should have excellent prior knowledge of how the instrument is performing. This knowledge can be used to set informative priors on the knot values of the splines.

\subsection{Averaged TDI transfer function}\label{sec:avg_TDI_response}

Equation~\eqref{eq:drift_equation} demonstrated that the zeros of the transfer functions drift due to (1) the segment length and (2) the evolution of the arm-lengths. We account for this slow drift by averaging the TDI transfer functions over a specific number of light travel times for each segment.

Take a segment of data with duration $T_{\rm seg}$ and divide it further into $n_e$ uniform segments with $t_k$ centred on each such that 
\begin{equation}\label{eq:t_k_equation}
t_k = t_0 + \frac{(k - 1/2)}{n_e}T_{\rm seg}, \qquad k \in \{1,2,\ldots, n_e\}\,.
\end{equation}
Over the full segment, we make the assumption that cross-correlations between the pieces are negligible. Similarly, since the evolution of the LTTs is extremely small over a piece, we further assume that each piece is locally stationary. We can then approximate 
\begin{align}\label{eq:split_up_expectation}
\mathbb{E}[\tilde{\boldsymbol{d}}\tilde{\boldsymbol{d}}^{\dagger}] &\approx \sum_{k}\mathbb{E}[\tilde{\boldsymbol{d}}_{k}\tilde{\boldsymbol{d}}_{k}^{\dagger}] \nonumber \\
& = \frac{N}{2\Delta t} \frac{1}{n_e}\sum_{k} \bM_k \bSigma^{(k)}_{\eta}\bM^{\dagger}_{k}\,.
\end{align}
The mixing matrix of each term in the sum uses the LTTs at the $t_k$ of Eq.~\eqref{eq:t_k_equation}.

By writing
\begin{align}
    \bar{\bK}_{\rm oms} &= \frac{1}{n_{e}}\sum_{k=1}^{n_{e}} \bM_k\bM_k^{\dagger}\,, \nonumber\\
    \bar{\bK}_{\rm tm} &= \frac{1}{n_{e}}\sum_{k=1}^{n_{e}} \bK_{\rm tm}\big(\bM_k,\varphi^{(k)}\big)\,, \label{eq:averaged_kernels}
\end{align}
and assuming that both $S_{\rm oms}$ and $S_{\rm tm}$ are constant in time over the full record of data, we can build the averaged instrumental covariance
\begin{empheq}[box=\fbox]{equation}\label{eq:avg_kernel_factorisation}
    \bar{\bSigma}_{\rm inst}(f) = \Stm^{\rm ff}(f)\,\bar{\bK}_{\rm tm}(f) + \Soms^{\rm ff}(f)\,\bar{\bK}_{\rm oms}(f)\,.
\end{empheq}

Notice that this averaging scheme is independent of the noise parameters, so only needs to be performed once at set-up. For $\Delta f = 1/T$ fixed, between consecutive epochs the LTTs change by $\delta \bar{d} / n_e$. With $n_e = 8$, Eq.~\eqref{eq:drift_equation} gives $\delta f_{k}/\Delta f = k$, so the zeros of consecutive epochs are one, two and three bins apart and the average fills the drift without gaps.

\begin{figure}[t]
    \centering
    \includegraphics[width=\linewidth]{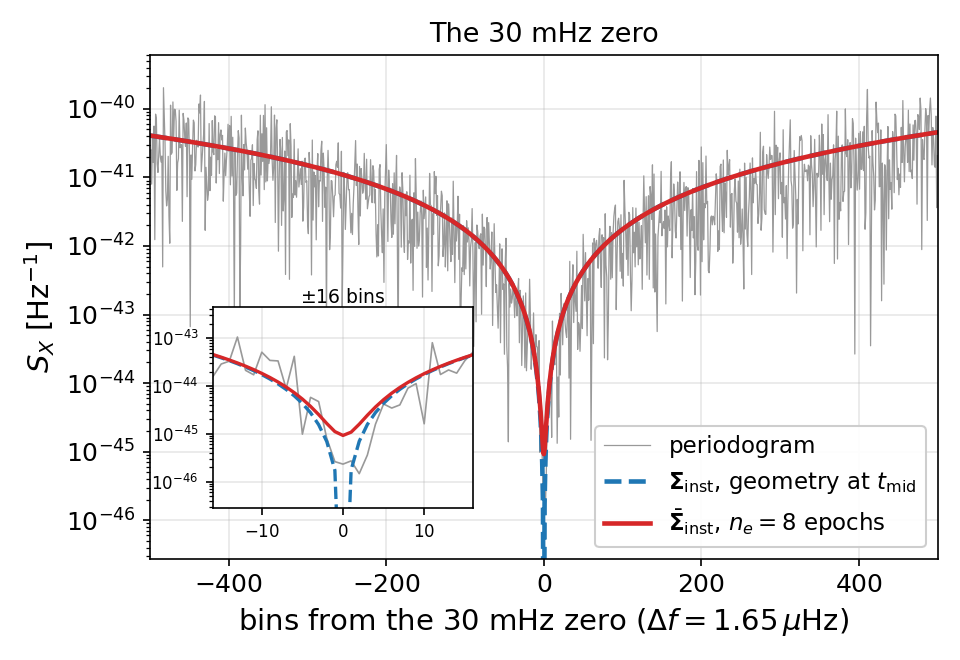}
    \caption{The \SI{30}{mHz} zero of $X$ on the first segment, $500$ bins to either side. The periodogram of the conditioned data (grey) against the instrument covariance at the injected spectra with the geometry frozen at the middle of the segment, $\bSigma_{\rm inst}$ (blue dashed), and with the kernels averaged over $n_e = 8$ epochs, $\bar{\bSigma}_{\rm inst}$ of Eq.~\eqref{eq:avg_kernel_factorisation} (red). The two differ only within a few bins of the zero, shown in the inset. The zero of the data drifts by eight bins over the week and is smeared over them. The frozen covariance falls to zero at one bin, the averaged one follows the data. Produced by \srclink{fig_conditioning.py}.}
    \label{fig:averaged_zero}
\end{figure}

Figure~\ref{fig:averaged_zero} compares the periodogram of the data at the first zero with the averaged instrument covariance with $n_e = 8$ and the static covariance with $t = t_{\rm mid}$. The periodogram is smeared over 8 bins, and the static covariance drops to zero at $f = 30\,$mHz, far below the data at that zero. The averaged covariance tracks the smeared zero. 

In Appendix~\ref{app:sec_checking_avg_transfer_func}, we have carefully checked that $n_{e} = 8$ is a sufficient number of epochs required to treat the time-evolving data. Results in that appendix are tabulated in Table~\ref{tab:geometry} and Fig.~\ref{fig:geometry_grid}. 

\subsection{The covariance of the conditioned data}\label{subsec:window}

The likelihood of Eq.~\eqref{eq:whittle} is a statement about the covariance of the data it is handed, and the data it is handed are not the instrument's noise process but that process filtered, decimated, segmented and tapered. Each of those operations is linear, and each changes the covariance in a way that can be written down exactly. 
Here we state the result with mathematical details relegated to Appendix~\ref{app:conditioning}. We approximate the covariance of the filtered, decimated and tapered data of a segment as
\begin{empheq}[box=\fbox]{align}
    \bar{\bSigma}_{\rm data}(f_j;\blambda) &= \frac{1}{N^{2}}\sum_{p} \big|\tilde{w}(f_j - f_p)\big|^{2} \nonumber\\
    &\qquad\times\big[R(f_p)\, \bar{\bSigma}_{\rm inst}(f_p;\blambda) + \bA(f_p;\blambda)\big]\,. \label{eq:data_model_cov}
\end{empheq}
Here $\bar{\bSigma}_{\rm inst}$ is the averaged instrument covariance of Eq.~\eqref{eq:avg_kernel_factorisation}. $R$ is the response of the two filters, the band-pass filter and the anti-alias filter of the downsampling. The matrix $\bA$ describes the power aliased by the decimation. Finally, $\tilde{\bw}$ is the DFT of the taper on the decimated grid. Equation~\eqref{eq:data_model_cov} replaces $\bSigma$ in the likelihood of Eq.~\eqref{eq:whittle}. There are two steps in our noise inference pipeline. The first is a post-processing (with respect to the instrument data) where the instrumental data are treated as in Sec.~\ref{subsec:simulation} using \texttt{MojitoProcessor}. The second is where the noise spectra are inferred, folding the post-processing steps into the noise covariance matrix. The filtering, decimation, windowing and segmentation procedures are known in advance (by the analyst) and are therefore not sampled over.

The settings are all recorded in the output file of \texttt{MojitoProcessor} and are folded into the averaged instrumental covariance matrix. Below, we discuss how the instrumental covariance is altered by each post-processing step. 

The Butterworth band-pass filter with filter response $H_{\text{bw}}(f)$ is applied forwards and backwards, which makes it zero phase. The anti-alias finite impulse response (FIR) $H_{\text{fir}}(f)$ of the decimation procedure is applied once. Both filters are linear time-invariant filters acting identically on every channel. Together they multiply the covariance by (see Eq.~\eqref{eq:lti_spectrum})
\begin{equation}\label{eq:response}
    R(f) = |H_{\rm bw}(f)|^{4}\,|H_{\rm fir}(f)|^{2}\,.
\end{equation}
Given our simulation parameters in Table~\ref{tab:sim_params}, the filter response is not flat\footnote{It is unity to a few parts in $10^{3}$ between \SI{0.3}{} and \SI{30}{mHz}; it falls to $0.886$ at \SI{0.1}{mHz} through the high-pass corner, and to $0.94$, $0.42$ and $0.19$ at \SI{60}{}, \SI{90}{} and \SI{100}{mHz} through the low-pass corner and the transition band of the FIR.} over the science band $f \in [10^{-4},10^{-1}]\,$Hz. Since it is not constant over the analysis band, the filter effects creep into the data and modify the noise process, requiring a correction to the covariance matrix describing the data.  

Keeping every $D$th sample of the filtered record folds its spectrum, in strips of width $f_s = 1/\Delta t$, onto the band of the decimated data. The one-sided cross-spectral matrix of the filtered, decimated data at $f \in [0, f_s/2]$ is
\begin{align}
    \bar{\bSigma}_{\rm filt}(f) &= \underbrace{R(f)\,\bar{\bSigma}_{\rm inst}(f)}_{\text{science band}} + \underbrace{\bA(f)}_{\text{aliased}}\,, \label{eq:alias_fold}\\
    \bA(f) &= \sum_{k = 1}^{D/2} \Big[R(f + k f_s)\,\bar{\bSigma}_{\rm inst}(f + k f_s) \nonumber\\
    &\qquad\quad + R(k f_s - f)\,\bar{\bSigma}_{\rm inst}(k f_s - f)^{*}\Big]\,. \nonumber
\end{align}
The first term is the band itself, multiplied by the filter response of Eq.~\eqref{eq:response}. The sum is the aliased power. Its first part is the strips above the band shifted down by $k f_s$, its second the strips mirrored about the Nyquist frequency, which enter conjugated because the time series is real. Each term is the instrument covariance at a source frequency above the band, suppressed by the response there. Appendix~\ref{app:decimation} derives Eq.~\eqref{eq:alias_fold} from the autocovariance of the decimated record and shows that, for the filter of Table~\ref{tab:sim_params}, only the first $k = 1$ mirrored order survives.

The matrix $\bA$ depends on $\blambda$ through the source frequencies. For the first mirrored order the source $f_s - f$ runs from \SI{0.2}{} down to \SI{0.1}{Hz} across the band. Over the top half of the band it lies between \SI{0.1}{} and \SI{0.15}{Hz}, within reach of the upper ghost knot of the splines, and that part of $\bA$ is carried as a fitted block. Sources beyond the ghost knot are evaluated with the seed shape and are fixed at set-up. 

The aliased power is negligible over most of the band and matters at the two upper zeros, \SI{60}{} and \SI{90}{mHz}. Here the instrument covariance falls some seven and four orders of magnitude below the conditioned data (Fig.~\ref{fig:data_vs_instrument}, bottom). This matters for sources with power at the top of the band, such as extreme mass-ratio inspirals, stellar-origin black hole binaries and the lightest massive black hole binaries. 

After filtering/decimation and segmentation, we apply a window to the data to mitigate spectral leakage. Windowing a stationary noise process induces a non-stationary feature, giving correlated noise bins in the frequency domain. For more information, see~\cite{Talbot:2021igi, Edy:2021par, Burke:2025bun, burke2026mindsecondgapimpact}.
In line with~\cite{burke2026mindsecondgapimpact}, we use its leading-diagonal form, Eq.~(91) of Ref.~\cite{burke2026mindsecondgapimpact}, treating the correlations between bins as subdominant
\begin{equation}\label{eq:windowed_diag}
    \bar{\bSigma}_{\text{data}}(f_j;\boldsymbol{\lambda}) = \frac{1}{N^2}\sum_{p} \big|\tilde{w}(f_j - f_p)\big|^{2}\,\bar{\bSigma}_{\rm filt}(f_p;\boldsymbol{\lambda})\,,
\end{equation}
the sum running over the full two-sided grid, with $\bar{\bSigma}_{\rm filt}(-f) = \bar{\bSigma}_{\rm filt}(f)^{*}$ entry-wise and the index $j - p$ taken modulo $N$. 
Including the diagonal correction to the covariance matrix arising from the window increases the computational complexity of the algorithm by $\mathcal{O}(p\cdot N\log N)$ for $p$ the number of retained bins in the analysis. However, with an analysis of gaps planned for future work, we keep the facility to fold an arbitrary window into the covariance. 

The convolution is evaluated exactly only where $\bar{\bSigma}_{\rm filt}$ varies over the width of the window. This is of the order of a few tens of bins, and is particularly important near the zeros of the transfer function (see Ref.~\cite{burke2026mindsecondgapimpact}) and the steep low-frequency edge of the band. Otherwise, we can use the scalar approximation to the covariance $\overline{w^{2}}\,\bar{\bSigma}_{\rm filt}$, Eq.~(99) of Ref.~\cite{burke2026mindsecondgapimpact}. With our settings, $\overline{w^{2}} = N^{-1}\sum_{n} w[n]^{2} = 1 - 5\alpha/8 = 0.969$ for the Tukey window. We identify these regions at set-up by convolving the seed covariance exactly and comparing. We can do this because we assume knowledge of the underlying noise spectra, which have a simple shape. In practice, if the noise statistics of the underlying secondary noises were unknown, then sensibly we would require the full convolution at all frequency bins to build the leading diagonal of the covariance matrix. This increases the cost of a likelihood call by a factor of six (Fig.~\ref{fig:cost}). 

\subsection{Verification} \label{subsec:verification}

In this section, we will describe our whitening procedure and metric to determine whether our conditioned covariance Eq.~\eqref{eq:data_model_cov} accurately describes the statistics of the noise process.  
By factoring the covariance of the conditioned data set at frequency bin $j$, $(\bar{\bSigma}_{\rm data})_{j} = (\boldsymbol{L}\boldsymbol{L}^{\dagger})_{j} \in \mathbb{C}^{3\times 3}$, we can whiten,
\begin{equation}
\tilde{\boldsymbol{z}}_j = \sqrt{\tfrac{2\Delta t}{N}}\;(\boldsymbol{L}^{-1})_{j}\tilde{\bd}_j \sim \mathcal{CN}(0, \bI)\,, \label{eq:whitened_z_j}
\end{equation}
where $\tilde{\boldsymbol{z}}_{j} \in \mathbb{C}^{3}$ holds the whitened residuals of the three channels at bin $j$, with real and imaginary parts jointly distributed as a standard complex normal. The three channels are now uncorrelated, and we can build the single scalar value $q_j$ at bin $j$
\begin{equation}
    q_j = \tilde{\boldsymbol{z}}_j^{\dagger}\tilde{\boldsymbol{z}}_j = \sum_{k=1}^{3}|\tilde{z}_{j,k}|^{2} \sim \Gamma(3,1)\,, \label{eq:whitened_gamma}
\end{equation}
so that $q_j$ is the sum of the squared moduli of three independent standard complex Gaussians, which is Gamma distributed with mean three and scale one. Equivalently $2 q_j \sim \chi^{2}_{6}$, the chi-square of the six real components of $\tilde{\boldsymbol{z}}_j$, each of variance $1/2$. Equation~\eqref{eq:whitened_gamma} is a convenient packed representation of the whitened residuals $\tilde{\boldsymbol{z}}_{j}$ and will be used frequently throughout this manuscript. 

\begin{figure}[t]
    \centering
    \includegraphics[width=\linewidth]{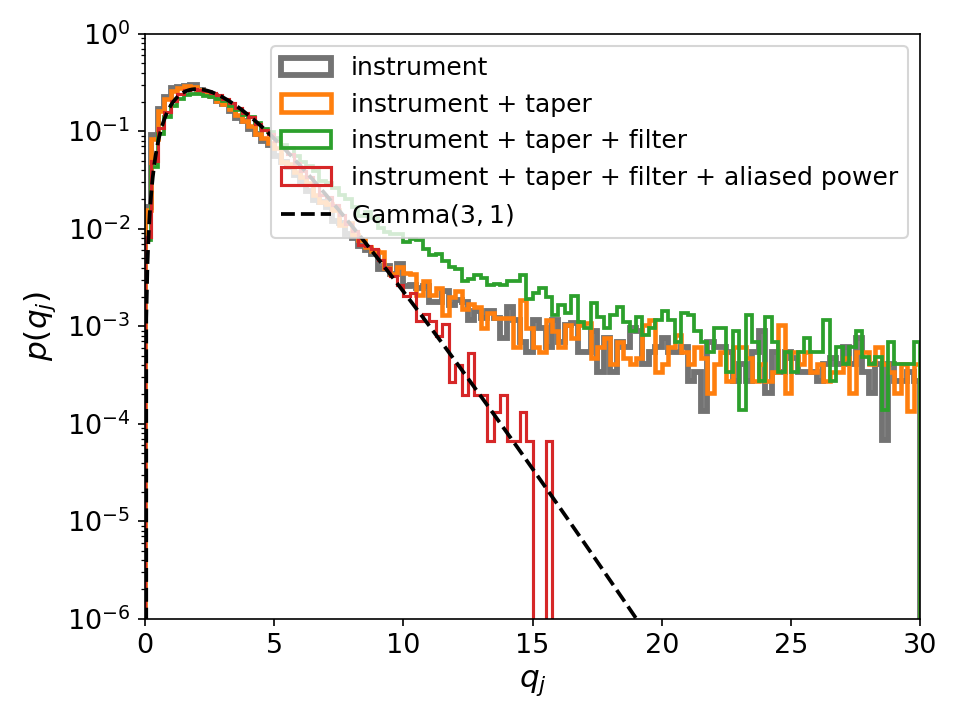}
    \caption{The distribution of $q_j$ over the $60\,420$ bins of the band for the four stages of the data model, the instrument covariance alone, with the taper, with the response and with the aliased power, which is Eq.~\eqref{eq:data_model_cov}, against the $\Gamma(3,1)$ law of Eq.~\eqref{eq:whitened_gamma}. The bulk is the same for every stage, since most of the band is unaffected by the conditioning. The three incomplete models leave the law at $q_j \simeq 8$ with a flat tail that runs off the axis, about $3\%$ of the bins; the full data model follows it to the last bin. Produced by \srclink{fig_stages.py}.}
    \label{fig:stages_hist}
\end{figure}

We now verify the effect each post-processing step has on the overall noise covariance matrix. We will compute $q_j$ in stages, first using the instrument covariance matrix, then incorporating the window, then filter and finally the aliased power leading to Eq.~\eqref{eq:data_model_cov}. 

\begin{figure}[t]
    \centering
    \includegraphics[width=\linewidth]{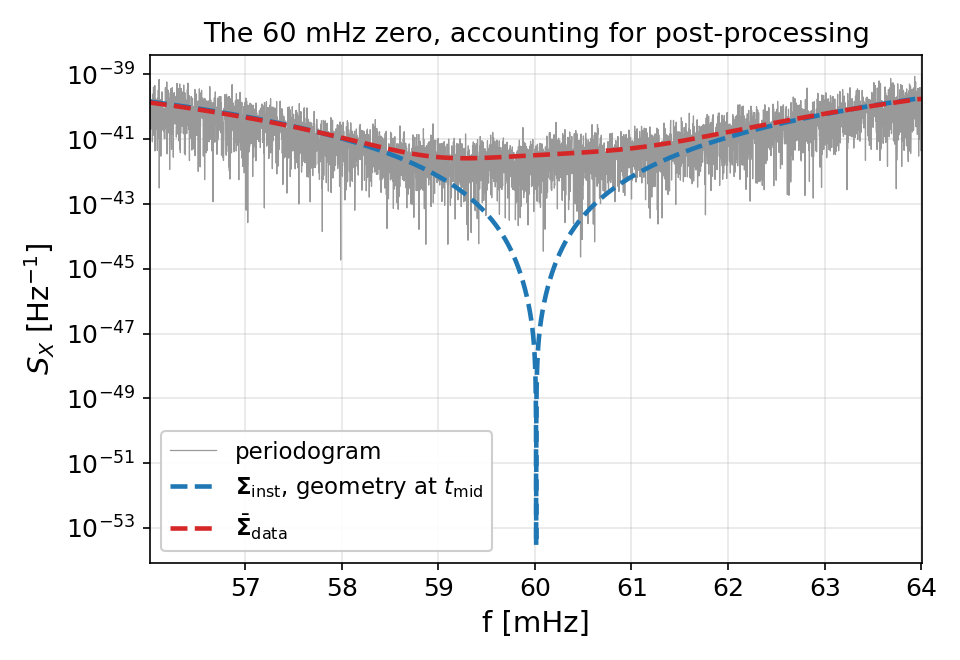}
    \caption{The \SI{60}{mHz} zero of $X$ on the first segment, as in Fig.~\ref{fig:data_vs_instrument} (bottom), with the data model added. The periodogram of the conditioned data (grey) against the instrument covariance $\bSigma_{\rm inst}$ with the geometry frozen at the middle of the segment (blue dashed) and the data model $\bar{\bSigma}_{\rm data}$ of Eq.~\eqref{eq:data_model_cov} (red dashed), both at the injected spectra. Produced by \srclink{fig_conditioning.py}.}
    \label{fig:data_model_zero}
\end{figure}

Figure~\ref{fig:stages_hist} plots the histogram of $q_j$ and compares it to the expected $\Gamma(3,1)$ distribution. Only when including the aliased power do the whitened residuals follow their expected distribution. This is due to the enormous values of $q_j$ around the zeros of the transfer function (see Fig.~\ref{fig:data_model_zero}), which is the reason for the very heavy tails in the mis-modelled covariance matrices. For the interested reader, the left-hand panel of Table~\ref{tab:stages} in Appendix~\ref{app:verification} gives the mean of $q_j$ over specific frequency bands whilst incorporating the post-processing effects incrementally. Figure~\ref{fig:data_model_zero} shows the result at the \SI{60}{mHz} zero. The data model is consistent with the periodogram of the data through the zero. 

\subsection{Likelihood and Inference schemes}\label{subsec:inference}
The likelihood is written up as an array backend that may be \texttt{numpy} or \texttt{jax.numpy}~\cite{harris2020array,jax2018github}. Since every array has a fixed shape JAX can perform a just in time (JIT) compilation of functions that are not sampled over (TDI transfer functions, filter responses etc.). All arithmetic is in double precision, JAX's 64-bit mode. With the likelihood written in \texttt{JAX}, the gradient of the log-likelihood with respect to the seventeen knot values can be computed at about three times the cost of a single likelihood call. Below we detail two inference schemes to estimate the noise spectra in order of increasing computational complexity. The first is maximum likelihood estimation (MLE) and the second is Bayesian inference.

Our MLE scheme makes use of the bound-constrained quasi-Newton optimiser, L-BFGS-B~\cite{Byrd1995lbfgsb,2020SciPy-NMeth}, which takes the analytic gradient of the likelihood, considers the prior box and iteratively converges to a maximum of the likelihood. 
A single start with tight priors around an assumed truth is the route a user would take if an expected noise curve were provided by experts monitoring the instrument on the ground. This is an excellent option with tight priors on a well understood instrument. With this method it is still possible to get stuck on local maxima of the likelihood, returning noise estimates inconsistent with the truth. 

We account for this by supplementing the MLE algorithm with a ``zoom'' scheme, which is robust to large uninformative priors on the knots. 
The scheme we use is a multi-start search that constrains the prior box as it progresses. In the first round $n_{s}$ (default $n_s = 100$) starting points are drawn uniformly over the chosen prior box. Each is optimised to the working tolerance $\sim 10^{-10}$. The $n_{k}$ (default $n_k = 3$) highest-likelihood points among those that converged become the centres of new boxes of half the width. In the next round we halve the number of starts and re-run the algorithm, obtaining the $n_k$ highest likelihood points and halving the boxes again. We perform $n_r$ (default $n_r = 3$) rounds, and the best found coordinate is used to seed a single-start MLE with a tolerance of $10^{-14}$. This is a polishing step. 
The result is a point estimate of the knot values at the maximum of the likelihood. The uncertainty can be estimated using the observed Fisher information matrix at the numerical maximum likelihood point $\blambda = \blambda_{\rm MLE}$
\begin{equation}\label{eq:laplace}
    \bGamma = -\frac{\partial^{2}\ln\mathcal{L}}{\partial\blambda\,\partial\blambda}\bigg|_{\blambda_{\rm MLE}}\,, \qquad {\rm Cov}\big[\blambda_{\rm MLE}\big] \simeq \bGamma^{-1}\,.
\end{equation}
We compute $\bGamma$ by central differences of the automatic gradient with a step of $10^{-4}$ in $\log_{10}$. The marginal uncertainty of knot $\alpha$ reported via Eq.~\eqref{eq:laplace} is denoted $\sigma_{{\rm MLE},\alpha} = \sqrt{[\bGamma^{-1}]_{\alpha\alpha}}$. 

With the MLE and associated uncertainty estimate, it is then possible to aid sampling algorithms as discussed below. This leads to the second route, full Bayesian inference, sampling the posterior by MCMC. 

We use the parallel-tempered ensemble sampler \texttt{Eryn}~\cite{Karnesis:2023ras,Foreman-Mackey:2012any}, $40$ walkers at each of $3$ temperatures, with priors set as $\pm 10$ in $\log_{10}$ of the seeded knot values. We run $500$ burn-in iterations, which are discarded, followed by $2\,000$ iterations that are kept. With our settings, we obtain $80\,000$ cold-chain samples from $296\,052$ likelihood evaluations in \SI{10}{min}. The cold chain accepts $24\%$ of its proposals. The likelihoods of the proposed walkers are evaluated using 8 threads, each through the same compiled likelihood call. For the initial starting values, we begin the walkers in a ball of the Fisher-based width of Eq.~\eqref{eq:laplace} about the zoom maximum to speed up the search for the global maximum. The priors are still set as $\pm 10$ in $\log_{10}$. The cold chain gives the posterior samples. The spectra of Sec.~\ref{subsec:results_mcmc} and their credible bands are evaluated on a thinned subset of it. 
 
\subsection{Computational cost} 
\begin{figure}[t]
    \centering
    \includegraphics[width=\linewidth]{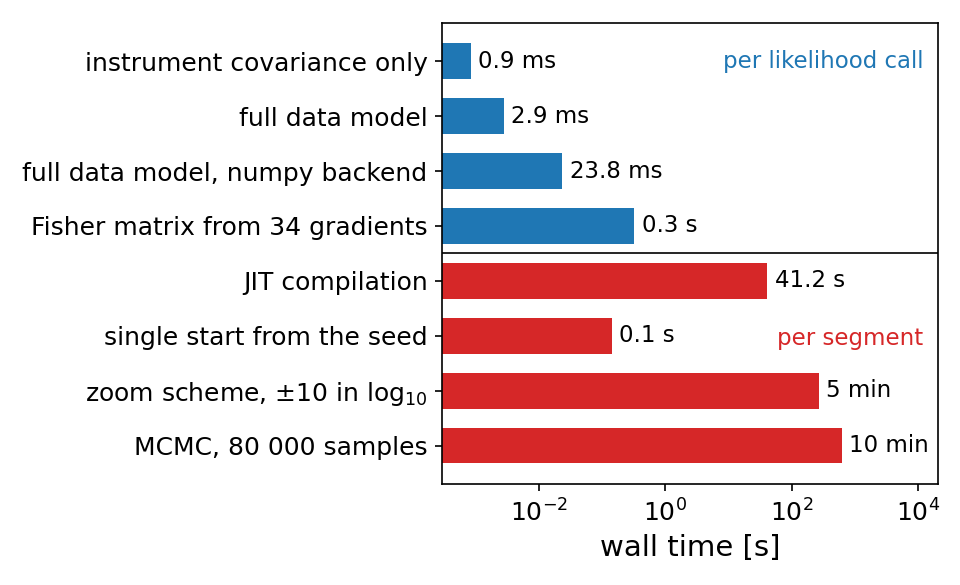}
    \caption{Cost of the pipeline for one seven-day segment ($N_f = 60\,420$ bins, $17$ parameters) on a laptop CPU (Apple M3 Max), one process with JAX's default threading. \emph{Blue}: one likelihood call with the instrument covariance alone and with the full data model of Eq.~\eqref{eq:data_model_cov}, the same call on the \texttt{numpy} backend, and the Fisher matrix from $34$ gradient evaluations. The likelihood numbers are medians over $50$ calls. \emph{Red}: the JIT compilation, which also builds the kernels, the aliased power and the convolution regions, once per segment; then the single start from the seed, the zoom scheme over the $\pm 10$ box and the sampler. Produced by \srclink{fig_cost.py} from \srclink{timings.py}, \srclink{run_mle.py} and \srclink{run_mcmc.py}.}
    \label{fig:cost}
\end{figure}
Figure~\ref{fig:cost} collects the timings with explanations in the caption.
Every timing in this paper comes from one machine. We have used a MacBook Pro with an Apple M3 Max, twelve performance and four efficiency cores and \SI{128}{GB} of memory, with Python~3.12, JAX~0.10 on its CPU backend, NumPy~2.4 and SciPy~1.17. No GPU or cluster was used to compute these timings. The numbers presented here are what a recent laptop (2026) delivers out of the box. We use the default threading that JAX would use on the CPU. 

\section{Results}\label{sec:results}

Everything in this section is obtained on the four seven-day segments of the simulation described in Sec.~\ref{subsec:simulation}. The full data stream is conditioned with the post-processing parameters of Table~\ref{tab:sim_params} and segmented into $T = 7$ day segments. Our analysis will refer to the first segment unless stated. All results come from one simulated month, so one realisation of the noise. The four segments are four independent realisations of the same process, and the twenty random starts of Fig.~\ref{fig:seed_mc} are independent fits of one of them. Across these, the fits agree within their widths and the covariance passes the same test in every segment (Fig.~\ref{fig:sweep}). A calibration over many realisations is left for future work. The covariance model is the full data model of Eq.~\eqref{eq:data_model_cov} with the geometry averaged over $n_e = 8$ epochs, and the knots of the TM and OMS spectra are those of Fig.~\ref{fig:model_ingredients}. 

\subsection{Maximum Likelihood Estimation}\label{subsec:results_zoom}

We begin by performing a single L-BFGS-B run from the parametric seed. With the automatic gradient of the \texttt{JAX} likelihood it converges in $10$ iterations and $15$ gradient evaluations. The set up is $\sim 41$ seconds to fully \texttt{JAX} compile the functions. The optimisation itself then takes \SI{0.14}{s} of wall time. Every one of the seventeen knots ends within $0.033$ in $\log_{10}$ of its injected value. 

\begin{figure}[t]
    \centering
    \includegraphics[width=\linewidth]{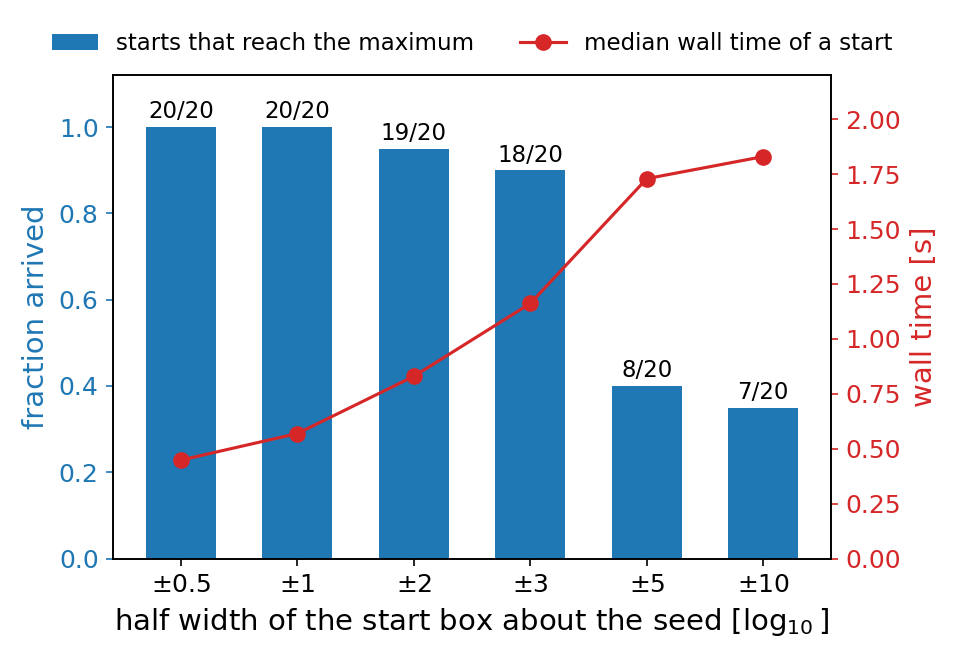}
    \caption{The single start from random seeds on segment 0. Twenty starts are drawn uniformly from a box about the seed, of the half width on the axis, and each is optimised with L-BFGS-B. Bars: the fraction that reach the maximum, with the count over each bar. Line: the median wall time of one start, after the \SI{41}{s} set-up. Within $\pm 3$ in $\log_{10}$ every start but three arrives; beyond that two thirds end on a local maximum. Produced by \srclink{fig_seed_mc.py}.}
    \label{fig:seed_mc}
\end{figure}

Figure~\ref{fig:seed_mc} runs the L-BFGS-B fit from starting points drawn progressively farther from the seed. Convergence to local maxima rather than the global maximum occurs for two reasons. The one failure at $\pm 2$ sits at the three lowest test-mass knots. The failures at $\pm 3$ and beyond sit at the test-mass knots between \SI{5}{} and \SI{30}{mHz}, where the test-mass and OMS spectra trade off. 
No run failed to converge. Every one of the $6 \times 20 = 120$ runs satisfied the optimiser's own test, and the number of iterations grew only from $39$ to $172$ across the whole range. What we learn from this figure is that the single-start scheme should only be used if exceptionally tight priors are placed on a known instrumental seed that is believed to be close to the truth. This scheme, although cheap, does risk converging onto local maxima. A single start appears reliable only within one in $\log_{10}$ of the truth, mostly reliable within two or three, and no further.

\begin{figure}[t]
    \centering
    \includegraphics[width=\linewidth]{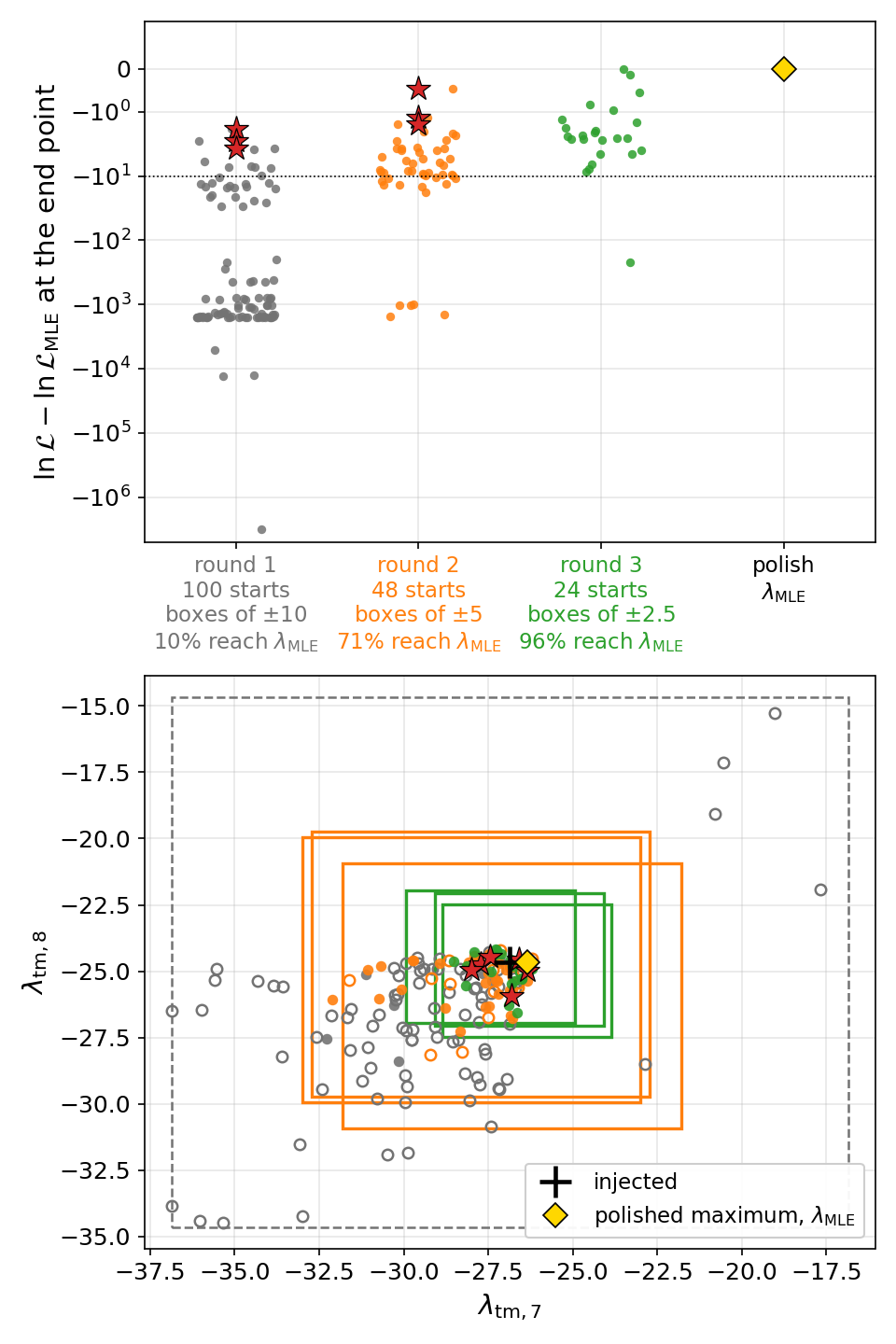}
    \caption{The zoom scheme on segment 0, prior box $\pm 10$ in $\log_{10}$ about the seed. \emph{Top}: the log-likelihood of every start after optimisation, relative to the polished maximum $\lambda_{\rm MLE}$ (gold diamond). The stars are the best three of a round, which become the centres of the boxes of the next round, of half the width. The dotted line is $10$ in log-likelihood below the maximum, the threshold of the percentages on the axis. \emph{Bottom}: the same rounds in the plane of the two least constrained knots, the test-mass knots at \SI{28}{} and \SI{100}{mHz}. The boxes of each round are drawn in its colour. Filled points are within $10$ in log-likelihood of the maximum, open points are further below. The spread of the filled points is the width of the likelihood in these knots, $1.3$ in $\log_{10}$ in Sec.~\ref{subsec:results_mcmc}. Produced by \srclink{fig_zoom.py}.}
    \label{fig:zoom}
\end{figure}

Figure~\ref{fig:zoom} depicts the zoom scheme in action, for the prior box of $\pm 10$ in $\log_{10}$ about the seed with explanations in the caption of the figure. The first round draws $100$ starts over the box. In the first, second and third rounds, $10\%$, $71\%$ and $96\%$ of the starts converge to the global maximum respectively. The coordinate of greatest likelihood is then refined using L-BFGS-B with a working tolerance of $\sim 10^{-14}$.

The total wall time including set up is about five minutes per segment, and the segments can be run in parallel. The uncertainty of the knot values is then estimated from Eq.~\eqref{eq:laplace}. The result of this MLE scheme feeds into the parameter sampling described below.

\begin{figure}[t]
    \centering
    \includegraphics[width=\linewidth]{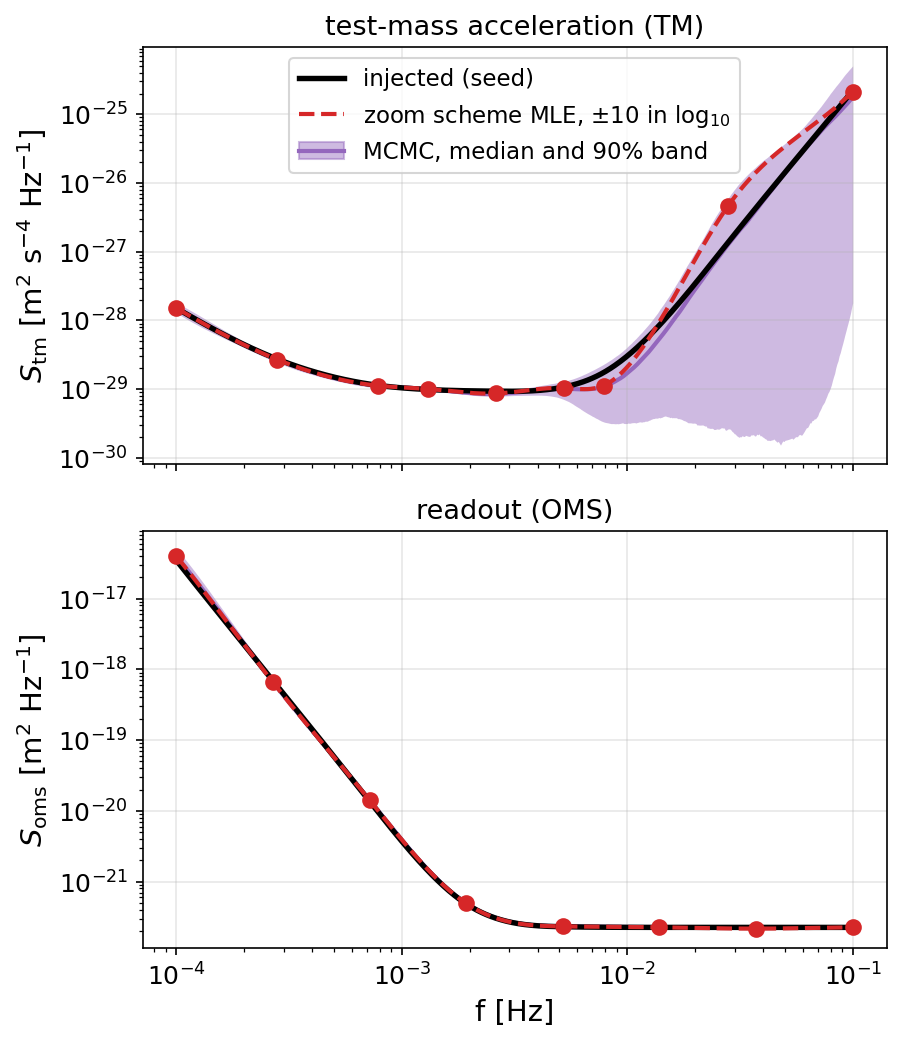}
    \caption{Maximum likelihood and the sampled posterior on the first segment. The recovered test-mass (top) and readout (bottom) spectra: the injected seed (black), the zoom maximum at $\pm 10$ in $\log_{10}$ (red dashed, its knots marked) and the posterior median and $90\%$ band of the chain (purple). Both follow the seed across the band. The posterior band opens downwards above \SI{8}{mHz} (Sec.~\ref{subsec:results_mcmc}). Produced by \srclink{fig_mle_mcmc.py}.}
    \label{fig:mle_spectra}
\end{figure}

\begin{figure}[t]
    \centering
    \includegraphics[width=\linewidth]{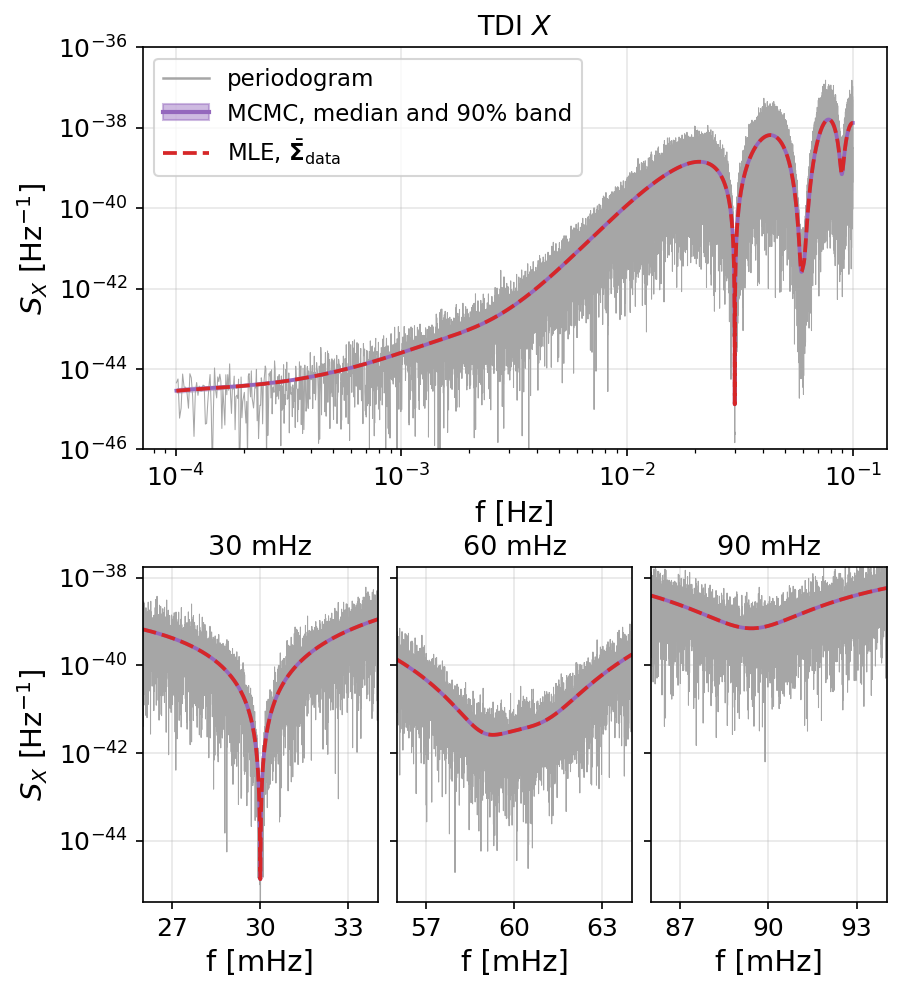}
    \caption{The $X$ channel on the first segment, over the band (top) and within \SI{4}{mHz} of each zero of the transfer (bottom). The periodogram of the conditioned data (grey), the data model $\bar{\bSigma}_{\rm data}$ of Eq.~\eqref{eq:data_model_cov} at the zoom maximum (red dashed) and its posterior median and $90\%$ band (purple). The bands are narrower than the lines except at the low edge of the band and at the bottoms of the zeros. Produced by \srclink{fig_mle_mcmc.py}.}
    \label{fig:mle_X}
\end{figure}

\subsection{Posterior sampling and the noise curves}\label{subsec:results_mcmc}

The marginalised posterior distributions are presented in Figs.~\ref{fig:corner_tm} and \ref{fig:corner_oms} with the results of the multi-start MLE scheme of Sec.~\ref{subsec:inference}. We also overlay a Gaussian of width $\sigma_{\rm MLE}$ on the posterior distributions. The MLE finds the knot values that describe the data and the Fisher-informed parameter uncertainty reasonably approximates the result of MCMC. Only for a subset of knot values, particularly the high frequency TM knots, do we see a breakdown in the Fisher approximation. This is to be expected: (1) the posterior distributions exhibit non-trivial skew (non-Gaussian) and (2) there is a clear degeneracy between OMS and acceleration noise at high frequency, where the two are highly correlated.

The recovered OMS and TM spectra with their $90\%$ credible intervals are shown in Fig.~\ref{fig:mle_spectra}. In Fig.~\ref{fig:mle_spectra}, we recover the knot values for both OMS and TM noise informed from the observed data. We constrain OMS well over the frequency band of interest. The TM noise, however, is well constrained only below \SI{10}{mHz}: above it the TM noise is a few per cent of the OMS noise, and the data bound its two knots from above only. The auto-spectrum of the conditioned noise of $X$ is shown in Fig.~\ref{fig:mle_X} alongside the result of the zoom scheme. The data model follows the periodogram over the band and through the zeros. 

In App.~\ref{app:real_imag_cross_spectra}, we analyse the results of the sampling specific to the cross spectra of the conditioned noise covariance matrix. In short, we accurately recover the real parts of the spectra and demonstrate via a Fisher-based formalism that the imaginary parts provide little enhancement of precision and no extra level of accuracy. We refer the interested reader to that appendix for more details. 

\begin{figure}[t]
    \centering
    \includegraphics[width=\linewidth]{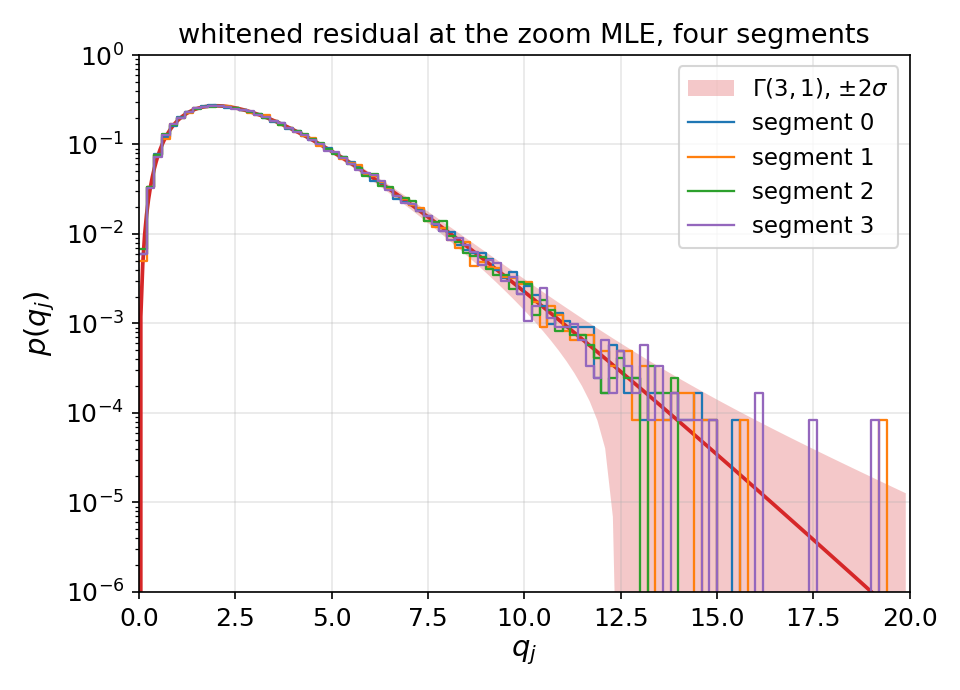}
    \caption{The fits across the thirty days. The histogram of the whitened residual $q_j$ of Eq.~\eqref{eq:whitened_gamma} at the zoom maximum of each of the four segments, against the $\Gamma(3,1)$ law (red line). The band is two standard deviations of the Poisson scatter of a histogram of $60\,420$ bins about the law. Produced by \srclink{fig_sweep.py}.}
    \label{fig:sweep}
\end{figure}

Figure~\ref{fig:sweep} includes the histogram of the whitened residual $q_j$ of Eq.~\eqref{eq:whitened_gamma} at the zoom maximum on the first segment, against the $\Gamma(3,1)$ law. Over all frequency bins the mean of $q_j$ is $2.999$ with standard error $\sqrt{3/n}\approx 0.007$. For the three zeros we obtain $3.3$, $2.7$ and $3.2$ over the $41$ bins respectively. The whitening also decorrelates neighbouring bins: the autocorrelation of $q_j$ at lags of one, two and ten bins is $0.007$, $-0.001$ and $-0.001$, so the bin-to-bin correlations that Eq.~\eqref{eq:windowed_diag} discards are negligible here. 

Finally, we analyse the full 30 days by segmenting the data into $T_{\rm seg} = 7$ day segments. Using the zoom scheme with a prior of $\pm 10$ in $\log_{10}$, we plot the whitened residuals $q_j$ of every segment in Fig.~\ref{fig:sweep}. Each segment follows the $\Gamma(3,1)$ law within the scatter of a histogram, demonstrating that the post-processed instrumental covariance matrix accurately describes the data. Each segment can be treated as independent, so the analysis could be run in parallel. The inference time of each segment is the same -- resulting in 5 minutes for a month, or two years' worth of data. 

\section{Conclusions and future work}\label{sec:conclusions}

\subsection{Summary}

We have presented and released a pipeline that estimates the instrumental noise of LISA from its second-generation TDI variables, in the frequency domain, on data conditioned as flight data will be. 

The noise of the six links needs two spectra, and the TDI variables need six light travel times. We represented the two fundamental noise spectra (OMS and TM) by cubic splines and propagated them through the mixing matrices of the delay polynomials at the light travel times of each segment. This provides the full $3\times 3$ cross-spectral matrix of the $\{X,Y,Z\}$ TDI variables, including correlations between the channels. Projecting the single-link variables through the mixing matrices produces the zeros of the TDI spectra, which our estimation routines fully account for. The result is an efficient likelihood, about $3$ milliseconds per call -- even accounting for the tapering, filtering and decimation post-processing applied to the data. Our analysis assumes that the light-travel times between arms are known exactly, and our pipeline uses them to construct the TDI transfer functions. The LTTs themselves drift within a segment due to the variation of the arm-lengths, and we treat this by averaging the transfer functions over the segment. This costs essentially nothing per likelihood call, and it is this feature that makes our noise model consistent throughout the zeros. 

The conditioning applied to the data should be modelled and thus accounted for during inference. The zero-phase filters multiply the covariance by a known response, the decimation folds the power above the new Nyquist frequency into the band, and the taper convolves the modelled spectra with its spectral window. Equation~\eqref{eq:data_model_cov} describes all three effects, in the order in which they are applied on the data. It is exact for the variance of every bin of a stationary record. With drifting arms it is the locally stationary approximation~\cite{Priestley1981,Dahlhaus1997}, tested via Table~\ref{tab:geometry} and Fig.~\ref{fig:geometry_grid}. Each effect has a region of the band where its absence is fatal for inference. With all three the whitened residual is consistent with a correct covariance in every band and through every zero, at about $2$\,ms per likelihood call above the instrument covariance alone.

A single gradient-based fit from the requirement curves recovers the injected spectra to within $0.03$ in $\log_{10}$ at every knot in a fraction of a second per seven-day segment, and a multi-start search over $\pm 10$ in $\log_{10}$ reaches the same maximum, to within one in log-likelihood, in five minutes. Against a parallel-tempered posterior sample, ten minutes on a laptop, the maximum agrees with the posterior mean to within $1.5$ posterior widths at every knot and to a fraction of one at fifteen, the posterior mean lies within two widths of the injected value at every knot, and the Fisher-based approximations to the posterior widths are the sampled ones to within $40$ per cent at the fifteen knots the data determine. The two they do not, the test-mass knots above \SI{10}{mHz} where that noise is a few per cent of the readout noise, are bounded from above only, as the sampler shows. Across the thirty-day simulation with unequal, time-varying arms, fitted segment by segment, the fits are consistent with each other and with the truth.

\subsection{Future work}

There are several extensions available for this work. The noise model of Eq.~\eqref{eq:link_cov} carries one test-mass spectrum and one readout spectrum shared by all six links. In more realistic scenarios, unequal noise levels across the constellation have been analysed in Refs.~\cite{Hartwig:2023pft,Kume:2024xvh,Vajpeyi:2026pzh,Liu:2026yzk}. The link covariance of Eq.~\eqref{eq:link_cov} could be generalised to twelve spectra, one per test mass and one per readout, modifying Eq.~\eqref{eq:kernel_factorisation} to instead have twelve kernels. For inference, the computational cost will scale with the number of parameters. This would be the natural next step for the work contained in this manuscript. 

The analysis performed here was conducted in the frequency domain. Other domains of interest, such as time-domain or time-frequency domain have their advantages to a problem of this kind. A time-domain inference would not need the same care against spectral artefacts, and may be the approach of choice for the short records of early alerts, where the settling time of the filters cannot be trimmed away. Similarly, the work here could be adapted to the time--frequency domain, using WDM methodologies~\cite{Cornish_2020,Digman:2022jmp,Vajpeyi:2026msr} or STFT~\cite{Speri_2026} representations. We believe the code is modular enough that only the likelihood would need to change. Conversion to the time-frequency domain would allow for the drifts of the zeros to be modelled directly as a time-domain quantity, rather than averaged over.   

The data here have no gaps, no glitches and no signals. Gaps could be treated through the window function as in Ref.~\cite{burke2026mindsecondgapimpact} and our codes should be able to handle records with missing data. Unfitted signals feature as a residual left in the data stream~\cite{Antonelli:2021vwg}, which could bias noise estimates. Similarly, short transient glitches or instrument malfunctions resulting in non-trivial time-dependent drifts of the noise covariance matrix impact noise estimation routines. The place where all of these matter at once is the global fit~\cite{Littenberg:2023xpl,Katz:2024oqg,GeeMoo,Strub:2024kbe}. Noise modelling is a crucial component within gravitational wave astronomy, arguably as important as GW signal analysis, so we wish to extend the \texttt{lisa-noise} package to account for all these extra features. 

\section{Data availability}

The noise inference code, \texttt{lisa-noise}, is archived at \href{https://doi.org/\lisanoisedoi}{DOI \lisanoisedoi}. Access to the code is restricted until the companion paper~\cite{Hoy:2026aaa} is released, when the record will be opened. The scripts that produce every figure and table of this manuscript, the results they were built from and the source of the manuscript are openly archived at \href{https://doi.org/\paperdoi}{DOI \paperdoi}. The script behind each figure and table is named in its caption. The conditioning package \texttt{MojitoProcessor}~\cite{Burke_MojitoProcessor_2026} is available on PyPI. The simulated data set can be regenerated with \texttt{lisainstrument}~\cite{bayle2023lisainstrument}, \texttt{lisaorbits}~\cite{Bayle_lisaorbits} and \texttt{pytdi}~\cite{Staab2022PyTDI} from the settings of Table~\ref{tab:sim_params}. The analysis further relies on \textsc{segwo}~\cite{Hartwig_segwo}, JAX~\cite{jax2018github} and \textsc{Eryn}~\cite{Karnesis:2023ras}.

\section{Acknowledgements}
We thank the developers of the \texttt{mojito} package and appreciate the useful discussions amongst many within the LISA Distributed Data Processing Centre. O.B.\ thanks the L01, L2D and SIM coordination units for useful discussions, in particular the noise subgroup of L2D, and Quentin Baghi, Nikos Karnesis, Federico Pozzoli and Martina Muratore for many fruitful conversations. O.B.\ appreciates comments made on an early draft by Avi Vajpeyi. O.B.\ also salutes John Veitch for the tip-off about how to treat the zeros of the TDI transfer functions by incorporating the post-processing into the covariance matrix. O.B. and J.B. acknowledge financial support from the Grant UKRI972 awarded via the UK Space Agency and acknowledge computational support from the \emph{andromeda} HPC at the University of Glasgow. C.H.\ thanks the University of Portsmouth for support through the Dennis Sciama Fellowship. L.K.N.\ thanks the UKRI Future Leaders Fellowship for support through the grant UKRI2745.

\appendix

\section{The mixing matrix and instrumental covariance}\label{app:mixing_matrix_inst_cov}

In this appendix, we write out the mixing matrix explicitly and derive the form of the instrument covariance matrix $\bSigma_{\rm inst}$. 

Following \textsc{segwo}, the $\eta$ variable ordering is 
\begin{equation}\label{eq:eta_vector}
    \tilde{\boldsymbol{\eta}} = \big(\tilde{\eta}_{12},\, \tilde{\eta}_{23},\, \tilde{\eta}_{31},\, \tilde{\eta}_{13},\, \tilde{\eta}_{32},\, \tilde{\eta}_{21}\big)^{\rm T}\,,
\end{equation}
and the components of the mixing matrix $\bM$ are
\begin{equation}\label{eq:mixing_explicit}
    \bM = \operatorname{diag}\big(1 - r_{1}r_{3},\; 1 - r_{1}r_{2},\; 1 - r_{2}r_{3}\big)\,\boldsymbol{N}\,,
\end{equation}
\begin{widetext}
\begin{equation}\label{eq:mixing_N}
    \boldsymbol{N} = \footnotesize
    \begin{pmatrix}
        1 - r_{3} & 0 & -z_{13}(1 - r_{1}) & -(1 - r_{1}) & 0 & z_{12}(1 - r_{3}) \\
        -z_{21}(1 - r_{2}) & 1 - r_{1} & 0 & 0 & z_{23}(1 - r_{1}) & -(1 - r_{2}) \\
        0 & -z_{32}(1 - r_{3}) & 1 - r_{2} & z_{31}(1 - r_{2}) & -(1 - r_{3}) & 0
    \end{pmatrix},
\end{equation}
\end{widetext}
with $r_{1} = z_{12}z_{21}$, $r_{2} = z_{23}z_{32}$, $r_{3} = z_{31}z_{13}$ dictating the \emph{zeros} of the TDI streams with $z_{ij} = \exp(-2\pi i f d_{ij})$, the delay convention of \textsc{segwo}. Here the rows should be understood as $X$, $Y$, $Z$ and the columns following Eq.~\eqref{eq:eta_vector}. 

The data are the TDI channels, $\tilde{\bd} = \bM\tilde{\boldsymbol{\eta}}$ [Eq.~\eqref{eq:mixing_def}], and $\bM$ is deterministic, so their covariance is that of the links projected through the mixing matrix,
\begin{equation*}
    \mathbb{E}\big[\tilde{\bd}\tilde{\bd}^{\dagger}\big] = \bM\,\mathbb{E}\big[\tilde{\boldsymbol{\eta}}\tilde{\boldsymbol{\eta}}^{\dagger}\big]\,\bM^{\dagger} = \frac{N}{2\Delta t}\,\bM\bSigma_{\eta}\bM^{\dagger}\,.
\end{equation*}
The matrix $\bM\bSigma_{\eta}\bM^{\dagger}$ above is the instrument covariance $\bSigma_{\rm inst}$ of Eq.~\eqref{eq:kernel_factorisation}. Our work primarily focuses on estimating the two spectra of Eq.~\eqref{eq:link_cov}, so the two can be separated via
\begin{equation*}
    \bSigma_{\eta} = \Soms^{\rm ff}\,\bI_{6} + \Stm^{\rm ff}\,\big(2\,\bI_{6} + \boldsymbol{\Phi}\big)\,,
\end{equation*}
with $\bI_{6}$  the $6\times 6$ identity and $\boldsymbol{\Phi}$ the matrix whose only non-zero entries are $\Phi_{ij,\,ji} = \varphi_{ij}$, one for each ordered pair of opposite links; it is Hermitian because $\varphi_{ji} = \varphi_{ij}^{*}$. Then
\begin{equation*}
    \bM\bSigma_{\eta}\bM^{\dagger} = \Soms^{\rm ff}\,\bM\bM^{\dagger} + \Stm^{\rm ff}\,\big(2\,\bM\bM^{\dagger} + \bM\boldsymbol{\Phi}\bM^{\dagger}\big)\,.
\end{equation*}
The last term has six contributions, two for each arm. With $\bM_{:,ij}$ the column of $\bM$ that multiplies $\tilde{\eta}_{ij}$,
\begin{align*}
    \bM\boldsymbol{\Phi}\bM^{\dagger} = \sum_{(ij,\,ji)} \Big(&\varphi_{ij}\,\bM_{:,ij}\,\bM_{:,ji}^{\dagger}\\
    &+ \varphi_{ij}^{*}\,\bM_{:,ji}\,\bM_{:,ij}^{\dagger}\Big)\,,
\end{align*}
the sum running over the three opposite-link pairs $(12,21)$, $(23,32)$, $(31,13)$, and the second term being the Hermitian conjugate (h.c.) of the first. Collecting terms by spectrum gives Eq.~\eqref{eq:kernel_factorisation} in the main text.

\section{Efficient computation of the multivariate Whittle likelihood}
\label{app:fast_inv_det}

By ordering the frequency-domain noise as
\begin{equation}
    \tilde{\bn} = \big(\tilde{n}^{(X)}_{0}, \tilde{n}^{(Y)}_{0}, \tilde{n}^{(Z)}_{0}, \tilde{n}^{(X)}_{1}, \dots, \tilde{n}^{(Z)}_{N_f-1}\big)^{\rm T} \in \mathbb{C}^{3N_f},
\end{equation}
the covariance of a stationary process is block diagonal, $\mathbb{E}[\tilde{\bn}\tilde{\bn}^{\dagger}] = \bigoplus_{k=0}^{N_f-1} \bSigma[k]$ with $\bSigma[k] \propto \bSigma(f_k)$ a $3\times 3$ Hermitian matrix, the independence of the Fourier components across frequency being what makes the off-diagonal blocks vanish. The determinant and the inverse of a direct sum are the direct sum of the determinants and inverses, $\ln\det\bigoplus_k \bSigma[k] = \sum_k \ln\det\bSigma[k]$ and $(\bigoplus_k \bSigma[k])^{-1} = \bigoplus_k \bSigma[k]^{-1}$, so both reduce to $3\times 3$ operations at every bin, and these are available in closed form. Writing
\begin{equation}
    \bSigma =
    \begin{pmatrix}
        C_{00} & C_{01} & C_{02} \\
        C_{01}^{*} & C_{11} & C_{12} \\
        C_{02}^{*} & C_{12}^{*} & C_{22}
    \end{pmatrix},
\end{equation}
with the indices $(0,1,2)$ standing for $(X,Y,Z)$ and every entry understood at the $k$th bin, the cofactors of the first row and of the diagonal are
\begin{align}
    a_{00} &= C_{11}C_{22} - |C_{12}|^{2}\,, &
    a_{01} &= C_{02}C_{12}^{*} - C_{01}C_{22}\,, \nonumber\\
    a_{02} &= C_{01}C_{12} - C_{02}C_{11}\,, &
    a_{11} &= C_{00}C_{22} - |C_{02}|^{2}\,, \nonumber\\
    a_{12} &= C_{02}C_{01}^{*} - C_{00}C_{12}\,, &
    a_{22} &= C_{00}C_{11} - |C_{01}|^{2}\,,
\end{align}
from which the determinant follows by expansion along the first row,
\begin{equation}\label{eq:det_formula}
    \det\bSigma = C_{00}\,a_{00} + C_{01}\,a_{01}^{*} + C_{02}\,a_{02}^{*}\,,
\end{equation}
real for a Hermitian matrix, and the inverse from the adjugate, $\bSigma^{-1} = {\rm adj}(\bSigma)/\det\bSigma$, with
\begin{equation}\label{eq:inverse_formula}
    {\rm adj}(\bSigma) =
    \begin{pmatrix}
        a_{00} & a_{01} & a_{02} \\
        a_{01}^{*} & a_{11} & a_{12} \\
        a_{02}^{*} & a_{12}^{*} & a_{22}
    \end{pmatrix}.
\end{equation}
Only six cofactors are needed for both, Hermiticity supplying the other three, and the six are shared between the determinant and the inverse. The point of writing them out is that they are elementwise expressions: storing the six unique entries of $\bSigma$ as arrays of length $N_f$, $\boldsymbol{c}_{ij} = (C_{ij}[0], \dots, C_{ij}[N_f-1])$, every cofactor, the determinant and the nine entries of the inverse are products and sums of those arrays, evaluated over all bins at once with no loop and no factorisation, in $\mathcal{O}(N_f)$ operations. The inverse is what the whitened residual of Eq.~\eqref{eq:whitened_gamma} uses.

The likelihood can be further optimised by pre-computing the data outer product. With $\boldsymbol{D}_k = \tilde{\bd}_k\tilde{\bd}_k^{\dagger}$, the outer product of the data at bin $k$, a Hermitian matrix fixed once the data are, the cyclic property of the trace gives
\begin{equation}\label{eq:quad_trace}
    \tilde{\bd}_k^{\dagger}\,\bSigma^{-1}[k]\,\tilde{\bd}_k = \tr\big(\bSigma^{-1}[k]\,\boldsymbol{D}_k\big) = \frac{\tr\big({\rm adj}(\bSigma[k])\,\boldsymbol{D}_k\big)}{\det\bSigma[k]}\,,
\end{equation}
and because both matrices are Hermitian the trace is real and involves only the six unique entries of each,
\begin{align}\label{eq:quad_cofactors}
    \tr\big({\rm adj}(\bSigma)\,\boldsymbol{D}\big) &= a_{00}\,|\tilde{d}_0|^{2} + a_{11}\,|\tilde{d}_1|^{2} + a_{22}\,|\tilde{d}_2|^{2} \nonumber\\
    &\quad + 2\,{\rm Re}\big(a_{01}D_{10} + a_{02}D_{20} + a_{12}D_{21}\big)\,,
\end{align}
with $D_{ji} = \tilde{d}_j\tilde{d}_i^{*}$ and every quantity at the $k$th bin. The six entries of $\boldsymbol{D}_k$ are computed once for all bins when the likelihood is set up. A call then evaluates the six cofactors, the determinant and Eq.~\eqref{eq:quad_cofactors}, three real and three complex products per bin for the quadratic form, where Eq.~\eqref{eq:whittle} taken literally would build the nine entries of the inverse and contract them twice with the data. The likelihood then takes the form
\begin{equation} 
    \ln\mathcal{L} \propto -\sum_k \ln\det\bSigma[k] - \frac{2\Delta t}{N}\sum_k \frac{\tr\big({\rm adj}(\bSigma[k])\,\boldsymbol{D}_k\big)}{\det\bSigma[k]}\,.
\end{equation} 

\section{The covariance of the conditioned data}
\label{app:conditioning}

This appendix derives the data model of Eq.~\eqref{eq:data_model_cov}, the response $R$ of Eq.~\eqref{eq:response} and the aliased power $\bA$ of Eq.~\eqref{eq:alias_fold}. The steps follow the order in which \texttt{MojitoProcessor} conditions the data. The filters act on the L1 record, the decimation on the filtered record and the taper on each segment of the decimated record. The taper is the case treated in Ref.~\cite{Burke:2025bun} and gives Eq.~\eqref{eq:windowed_diag}.

The instrument process is a real, zero-mean, stationary Gaussian vector process sampled at the L1 cadence $\Delta t_1 = 1/f_1 = 0.25$ seconds. Its one-sided cross-spectral matrix is $\bSigma_{\rm inst}(f)$ of Eq.~\eqref{eq:kernel_factorisation}. The pipeline uses the averaged $\bar{\bSigma}_{\rm inst}$ of Eq.~\eqref{eq:avg_kernel_factorisation} in its place, and nothing below depends on the distinction. It is convenient to work with two-sided spectra until the last step. For a real vector sequence $\bx[i]$ with autocovariance $\boldsymbol{\gamma}[m] = \mathbb{E}\big[\bx[i+m]\,\bx[i]^{\rm T}\big]$, the two-sided cross-spectral matrix $\boldsymbol{P}(f)$ on $[-f_1/2, f_1/2]$ is defined by
\begin{equation}\label{eq:autocov_def}
    \boldsymbol{\gamma}[m] = \int_{-f_1/2}^{f_1/2} \boldsymbol{P}(f)\,e^{2\pi i f m\Delta t_1}\,{\rm d}f\,.
\end{equation}
Because the sequence is real, $\boldsymbol{P}(-f) = \boldsymbol{P}(f)^{*}$, the entry-wise conjugate. The one-sided matrix counts the positive frequencies only, $\bSigma_{\rm inst}(f) = 2\boldsymbol{P}(f)$ for $f > 0$. We write
\begin{equation}\label{eq:breve_C}
    \breve{\bSigma}_{\rm inst}(g) \equiv \begin{cases} \bSigma_{\rm inst}(g)\,, & g > 0\,,\\ \bSigma_{\rm inst}(|g|)^{*}\,, & g < 0\,,\end{cases}
\end{equation}
for the one-sided matrix continued to negative frequencies, so that $\boldsymbol{P}(f) = \tfrac{1}{2}\breve{\bSigma}_{\rm inst}(f)$.

\subsection{Filtering}\label{app:filtering}

A filter with impulse response $h[a]$ acts on every channel as the convolution $\boldsymbol{y}[i] = \sum_a h[a]\,\bx[i - a]$. In the frequency domain the convolution is a product, $\tilde{\boldsymbol{y}}(f) = H(f)\,\tilde{\bx}(f)$, with $H(f) = \sum_a h[a]\,e^{-2\pi i f a\Delta t_1}$ the transfer function. The same scalar multiplies all three channels, so
\begin{equation}\label{eq:lti_spectrum}
    \mathbb{E}\big[\tilde{\boldsymbol{y}}\tilde{\boldsymbol{y}}^{\dagger}\big] = H\,\mathbb{E}\big[\tilde{\bx}\tilde{\bx}^{\dagger}\big]\,H^{*}\,, \qquad \boldsymbol{P}_y(f) = |H(f)|^{2}\,\boldsymbol{P}(f)\,.
\end{equation}
The cross-spectra are scaled exactly as the auto-spectra, and the phase of $H$ drops out.

Two filters act on the data. The Butterworth (bw) band-pass of Table~\ref{tab:sim_params} and the anti-aliasing filter through the downsampling routine. The band-pass filter is applied forwards and backwards over the reversed record, which makes it zero-phase. The backward pass has the transfer function $H_{\rm bw}^{*}$, so the pair applies $|H_{\rm bw}|^{2}$ to the data, with zero phase, and by Eq.~\eqref{eq:lti_spectrum} $|H_{\rm bw}|^{4}$ to the covariance. 

The anti-alias FIR is applied once, so it contributes $|H_{\rm fir}|^{2}$. Together,
\begin{equation}\label{eq:response_twosided}
    \boldsymbol{P}_{\rm filt}(f) = R(|f|)\,\boldsymbol{P}(f) = \tfrac{1}{2}\,R(|f|)\,\breve{\bSigma}_{\rm inst}(f)\,,
\end{equation}
with $R = |H_{\rm bw}|^{4}\,|H_{\rm fir}|^{2}$ the response of Eq.~\eqref{eq:response}. It is real, positive, even in $f$ and the same for every channel. Equation~\eqref{eq:lti_spectrum} assumes a filter that has been running on a process with infinite duration. This is the purpose of trimming the data at either end after applying the filter to remove the start-up transients. 

\subsection{Decimation}\label{app:decimation}

Keeping every $D$th sample of the filtered record is the one step that moves power across frequencies. 

The decimated process $\bx_d[m] = \boldsymbol{y}[mD]$ has cadence $\Delta t = D\Delta t_1$ and sampling rate $f_s = f_1/D$. In our simulation in Sec.~\ref{subsec:simulation}, we set $D = 20$ enabling a downsampled cadence of $\Delta t = 5\,$ seconds. The autocovariance of the decimated process is 
\begin{align}
\boldsymbol{\gamma}_d[m] &= \boldsymbol{\gamma}_{y}[mD] \nonumber \\
& = \int_{-f_1/2}^{f_1/2} \boldsymbol{P}_{\text{filt}}(f)\,e^{2\pi i f m D\Delta t_1}\,{\rm d}f \nonumber \\ 
& = \int_{-f_1/2}^{f_1/2} \boldsymbol{P}_{\text{filt}}(f)\,e^{2\pi i f m \Delta t}\,{\rm d}f\,. \label{eq:decimated_autocov}
\end{align}
Here we have used $D\Delta t_{1} = \Delta t$. Notice that we are assuming we have already applied both the band-pass and anti-aliasing filter to the data, hence the appearance of $\boldsymbol{P}_{\text{filt}}$. The filter acts over the frequencies $f \in (-f_1/2, f_1/2)$ and so contributes to the auto-covariance of the decimated data. We now describe how the power folds into the auto-covariance of the decimated process.

We can split the interval $[-f_1/2, f_1/2]$ up into strips of width $f_s$, with the $k$th strip being $[k f_s - f_s/2,\, k f_s + f_s/2)$ for $k \in \{-D/2, -D/2 + 1, \ldots, D/2\}$. The two end strips reach half a strip beyond the interval, where $\boldsymbol{P}_{\rm filt}$ vanishes. For $k = 0$, we find the strip of frequencies resolvable by assuming a cadence of $\Delta t$. Splitting the limits of Eq.~\eqref{eq:decimated_autocov}, one obtains 
\begin{equation}
\boldsymbol{\gamma}_{d}[m] =  \sum_{k=-D/2}^{D/2}\int_{k f_s - f_s/2}^{k f_s + f_s/2} \boldsymbol{P}_{\rm filt}(f)\,e^{2\pi i f m\Delta t}\,{\rm d}f\,.
\end{equation}
Now consider a change of variables $f = f' + kf_s$ with $f' \in [-f_s/2, f_s/2]$. We obtain
\begin{align}
\boldsymbol{\gamma}_{d}[m] &= \sum_{k=-D/2}^{D/2}\int_{-f_s/2}^{f_s/2} \boldsymbol{P}_{\rm filt}(f' + kf_s)\,e^{2\pi i (f' + kf_s) m\Delta t}{\rm d}f' \nonumber \\ 
& = \int_{-f_s/2}^{f_s/2} \left[\sum_{k=-D/2}^{D/2} \boldsymbol{P}_{\rm filt}(f' + k f_s)\right] e^{2\pi i f' m\Delta t}\,{\rm d}f'\,, \label{eq:alias_fold_twosided} 
\end{align}
Here we have used the fact that $e^{2\pi i (f' + k f_s) m\Delta t} = e^{2\pi i f' m\Delta t}$ since $f_s \Delta t = 1$ and $km\in\mathbb{Z}$. 

Within the end strips only the sources with $|f' + k f_s| < f_1/2$ contribute, since $\boldsymbol{P}_{\rm filt}$ vanishes beyond the L1 Nyquist frequency. Comparing with Eq.~\eqref{eq:autocov_def} on the coarse grid, the bracketed term is the two-sided spectral matrix of the decimated data,
\begin{equation}\label{eq:fold_twosided}
    \boldsymbol{P}_{d}(f) = \sum_{k=-D/2}^{D/2} \boldsymbol{P}_{\rm filt}(f + k f_s)\,, \qquad f \in [-f_s/2, f_s/2)\,.
\end{equation}

For $f_1 = \SI{4}{Hz}$ and $f_s = \SI{0.2}{Hz}$ that is $D = 20$, the band itself at $k = 0$ and nineteen alias orders. Restricting to the positive spectrum with  $f \in [0, f_s/2]$, we have $\bSigma_{\rm filt}(f) = 2\boldsymbol{P}_{d}(f)$. Inserting Eq.~\eqref{eq:response_twosided} for every strip,
\begin{equation}\label{eq:fold_onesided}
    \bSigma_{\rm filt}(f) = \sum_{k=-D/2}^{D/2-1} R\big(|f + k f_s|\big)\,\breve{\bSigma}_{\rm inst}(f + k f_s)\,.
\end{equation}
The $k = 0$ term is the band itself. The rest is the aliased power, which we denote $\bA$. For $k > 0$ the source $f + k f_s$ is positive and $\breve{\bSigma}_{\rm inst}$ is $\bSigma_{\rm inst}$ itself. For $k < 0$ the source is negative and, by Eq.~\eqref{eq:breve_C}, $\breve{\bSigma}_{\rm inst}$ is the conjugate of $\bSigma_{\rm inst}$ at $|k| f_s - f$, the strip mirrored about the Nyquist frequency. Pairing the orders $\pm k$ for $k = 1, \ldots, D/2$, we obtain 
\begin{align}
    \bA(f) = \sum_{k = 1}^{D/2} \Big[ &R(f + k f_s)\,\bSigma_{\rm inst}(f + k f_s) \nonumber\\
    &+ R(k f_s - f)\,\bSigma_{\rm inst}(k f_s - f)^{*} \Big]\,, \label{eq:alias_explicit}
\end{align}
with $D/2 = f_1/(2 f_s) = 10$ and the conjugate taken entry by entry. The direct term of the last order lies above $f_1/2$ and vanishes.

For the filter of Table~\ref{tab:sim_params} only the $k=1$ term survives. The response at every source beyond $f_s = \SI{0.2}{Hz}$ is below $10^{-28}$. What remains is the mirrored $k = 1$ term, i.e., the strip between \SI{0.1}{} and \SI{0.2}{Hz} folded about the Nyquist frequency. Defining $g = f_s - f$, we can write the aliased matrix
\begin{equation}\label{eq:alias_matrix}
    \bA(f) \simeq R(g)
    \begin{pmatrix}
        \Sigma_{XX}(g) & \Sigma_{XY}(g)^{*} & \Sigma_{XZ}(g)^{*} \\
        \Sigma_{XY}(g) & \Sigma_{YY}(g) & \Sigma_{YZ}(g)^{*} \\
        \Sigma_{XZ}(g) & \Sigma_{YZ}(g) & \Sigma_{ZZ}(g)
    \end{pmatrix},
\end{equation}
with $\Sigma_{XY} \equiv [\bSigma_{\rm inst}]_{XY}$. 

\section{Supplementary verification material}\label{app:verification}\label{app:sec_checking_avg_transfer_func}

\begin{figure}[t]
    \centering
    \includegraphics[width=\linewidth]{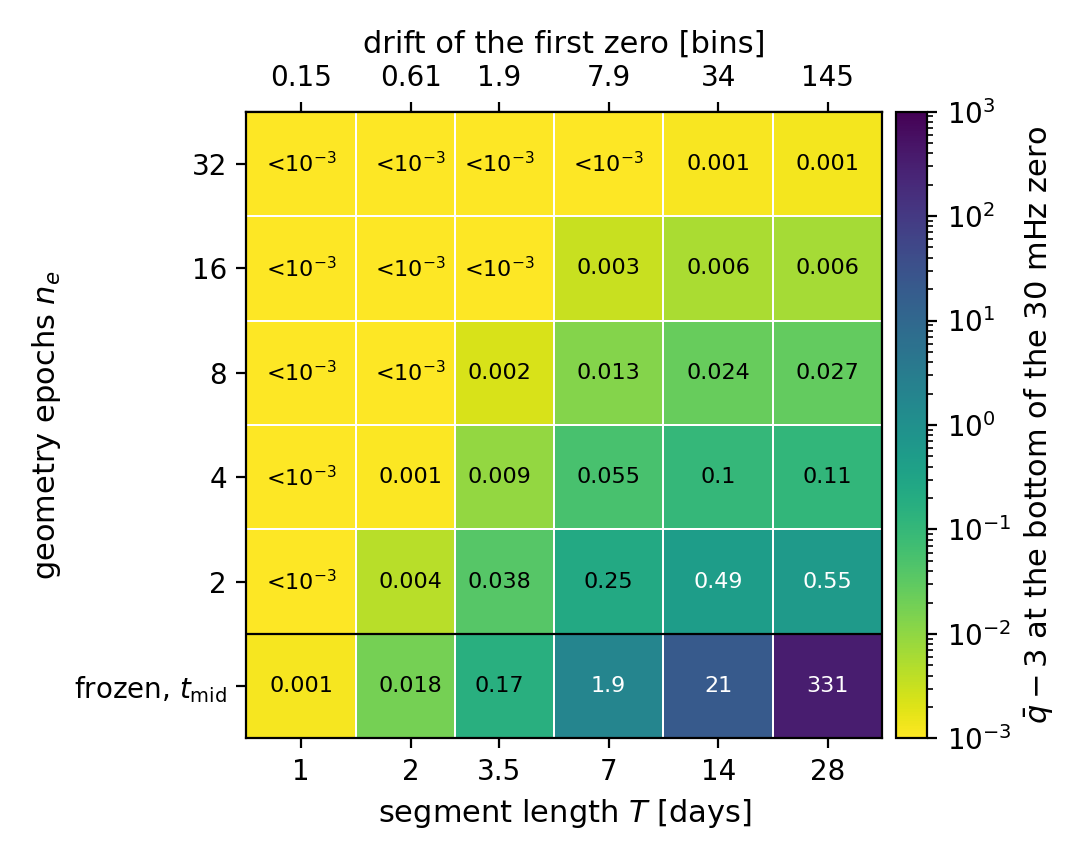}
    \caption{The colour is the mean whitened residual $\bar{q}$ of Eq.~\eqref{eq:whitened_gamma} over the seven bins at the bottom of the \SI{30}{mHz} zero, minus three, for the geometry frozen at $t_{\rm mid}$ (bottom row) or averaged over $n_e$ epochs, for segments of one to twenty-eight days. Along the top, the number of bins the zero drifts over the segment, which grows as $T^{2}$ by Eq.~\eqref{eq:drift_equation}. The frozen geometry is sufficient only for segments of about a day. For longer segments the averaged kernels are needed, and $n_e = 8$ suffices up to a month. We use $n_e = 8$ and $T = 7$ days for convenience. Nothing in the model prevents a single twenty-eight day segment. Produced by \srclink{geometry_grid.py}.}
    \label{fig:geometry_grid}
\end{figure}

This appendix collects further checks of the data model referred to in Secs.~\ref{sec:avg_TDI_response} and \ref{subsec:verification}. In Table~\ref{tab:geometry}, we compute the mean of $q_j$ with the full data model of Eq.~\eqref{eq:data_model_cov} to test the averaging procedure. The first column uses a covariance with the LTTs evaluated at the single instant $t = t_{\rm mid}$. The other columns average the kernels over $n_e$ epochs. Figure~\ref{fig:geometry_grid} repeats the test as the segment duration grows. Finally, Table~\ref{tab:stages} tests the data model stage by stage, adding the windowing, filtering and decimation one at a time to the averaged covariance matrix. The captions give more details. The mean of $q_j$ over a band is three when the covariance is consistent with the data. 

\begin{table*}[t]
\centering
\caption{Mean whitened residual $q_j$ of Eq.~\eqref{eq:whitened_gamma} at the injected spectra on segment 0, with only the geometry of the kernels of Eq.~\eqref{eq:averaged_kernels} varied between the columns. Bands and zeros as in Table~\ref{tab:stages}. The last column is the standard error of the band mean. Away from the zeros every column is consistent with $3$, so the geometry matters only at the zeros. At the \SI{30}{mHz} zero the frozen geometry leaves $4.1$ and costs $10$ in log-likelihood. Averaging over $n_e \ge 6$ epochs brings the residual to $3.3$ and the log-likelihood to within $0.03$ of $n_e = 16$. At the \SI{60}{} and \SI{90}{mHz} zeros the aliased power dominates and the geometry makes no difference. Below the residuals, the log-likelihood of the injected spectra relative to $n_e = 16$, the set-up time of the model, which evaluates the kernels once per epoch and grows with $n_e$, and the cost of a likelihood call, which does not depend on $n_e$. Produced by \srclink{geometry_table.py}.}
\label{tab:geometry}
\footnotesize
\setlength{\tabcolsep}{5pt}
\begin{tabular}{@{}lrrrrrrrr@{}}
\toprule
 & frozen & \multicolumn{6}{c}{averaged over $n_e$ epochs} & \\
\cmidrule(lr){2-2}\cmidrule(lr){3-8}
band [Hz] & $t_{\rm mid}$ & $2$ & $4$ & $6$ & $8$ & $10$ & $16$ & std.\ error \\
\midrule
$10^{-4}$--$10^{-3}$ & 2.99 & 2.99 & 2.99 & 2.99 & 2.99 & 2.99 & 2.99 & 0.07 \\
$10^{-3}$--$10^{-2}$ & 3.01 & 3.01 & 3.01 & 3.01 & 3.01 & 3.01 & 3.01 & 0.02 \\
$10^{-2}$--$3\times10^{-2}$ & 3.00 & 3.00 & 3.00 & 3.00 & 3.00 & 3.00 & 3.00 & 0.02 \\
$3$--$5\times10^{-2}$ & 3.00 & 2.99 & 2.99 & 2.99 & 2.99 & 2.99 & 2.99 & 0.02 \\
$5$--$6\times10^{-2}$ & 2.97 & 2.97 & 2.97 & 2.97 & 2.97 & 2.97 & 2.97 & 0.02 \\
$6$--$7\times10^{-2}$ & 3.02 & 3.02 & 3.02 & 3.02 & 3.02 & 3.02 & 3.02 & 0.02 \\
$7$--$8\times10^{-2}$ & 2.99 & 2.99 & 2.99 & 2.99 & 2.99 & 2.99 & 2.99 & 0.02 \\
$8$--$9\times10^{-2}$ & 2.98 & 2.98 & 2.98 & 2.98 & 2.98 & 2.98 & 2.98 & 0.02 \\
$9\times10^{-2}$--$10^{-1}$ & 3.02 & 3.02 & 3.02 & 3.02 & 3.02 & 3.02 & 3.02 & 0.02 \\
\midrule
\SI{30}{mHz} zero, $\pm 20$ bins & 4.07 & 3.42 & 3.35 & 3.33 & 3.33 & 3.33 & 3.33 & 0.27 \\
\SI{60}{mHz} zero, $\pm 20$ bins & 2.66 & 2.66 & 2.66 & 2.66 & 2.66 & 2.66 & 2.66 & 0.27 \\
\SI{90}{mHz} zero, $\pm 20$ bins & 3.21 & 3.21 & 3.21 & 3.21 & 3.21 & 3.21 & 3.21 & 0.27 \\
\midrule
$\Delta\ln\mathcal{L}$ at the injected spectra & $-10.4$ & $+0.15$ & $+0.08$ & $+0.03$ & $+0.02$ & $+0.01$ & $0$ & \\
set-up [s] & 7 & 10 & 18 & 25 & 33 & 41 & 64 & \\
likelihood call [ms] & 2.8 & 2.9 & 2.9 & 2.9 & 3.0 & 2.9 & 3.0 & \\
\bottomrule
\end{tabular}
\end{table*}

\begin{table*}[t]
\centering
\caption{Mean whitened residual $q_j$ of Eq.~\eqref{eq:whitened_gamma} at the injected spectra on segment 0, per band and per stage of the data model. The stages are the instrument covariance alone, $\bar{\bSigma}_{\rm inst}$, with the geometry averaged but nothing of the conditioning; the taper convolution, $|\tilde{\bw}|^{2}\ast\bar{\bSigma}_{\rm inst}$; the response added, $R\,(|\tilde{\bw}|^{2}\ast\bar{\bSigma}_{\rm inst})$, applied after the convolution, which differs from the order of Eq.~\eqref{eq:data_model_cov} only within a window width of the band edges; and the aliased power added, which is the full data model of Eq.~\eqref{eq:data_model_cov}. The left block is the data set of Table~\ref{tab:sim_params}, conditioned with the $\beta = 31$ anti-alias filter. The right block is the same record conditioned with $\beta = 5$. The last column is the standard error of the band mean, $\sqrt{3/n}$ for the $n$ bins of the band. The last three rows take the mean over the $41$ bins centred on each zero of the transfer. Reading across, the instrument covariance alone sits three per cent low below \SI{50}{mHz}, which is the scale of the taper. The taper convolution removes that. The response fixes the band between \SI{70}{} and \SI{80}{mHz} and raises the residual above it, where the fold adds the power that the response takes away. The aliased power brings every band and every zero to $3$. At $\beta = 5$ the aliased power sits at the bottom of the band instead, five orders of magnitude in the lowest decade, and again only the full model recovers $3$. Produced by \srclink{stages.py}.}
\label{tab:stages}
\footnotesize
\setlength{\tabcolsep}{4.5pt}
\begin{tabular}{@{}lrrrrrrrrr@{}}
\toprule
 & \multicolumn{4}{c}{$\beta = 31$} & \multicolumn{4}{c}{$\beta = 5$} & \\
\cmidrule(lr){2-5}\cmidrule(lr){6-9}
band [Hz] & $\bar{\bSigma}_{\rm inst}$ & $|\tilde{\bw}|^{2}\ast\bar{\bSigma}_{\rm inst}$ & $+R$ & $+\bA$ & $\bar{\bSigma}_{\rm inst}$ & $|\tilde{\bw}|^{2}\ast\bar{\bSigma}_{\rm inst}$ & $+R$ & $+\bA$ & std.\ error \\
\midrule
$10^{-4}$--$10^{-1}$, all bins & $8.7\times10^{4}$ & $6.5\times10^{4}$ & $6.9\times10^{4}$ & 3.00 & $1.1\times10^{3}$ & $1.1\times10^{3}$ & $1.2\times10^{3}$ & 2.99 & 0.007 \\
\midrule
$10^{-4}$--$10^{-3}$ & 2.88 & 2.97 & 2.99 & 2.99 & $1.2\times10^{5}$ & $1.2\times10^{5}$ & $1.3\times10^{5}$ & 2.94 & 0.07 \\
$10^{-3}$--$10^{-2}$ & 2.91 & 3.01 & 3.01 & 3.01 & 11.9 & 12.3 & 12.3 & 3.00 & 0.02 \\
$10^{-2}$--$3\times10^{-2}$ & 2.90 & 3.00 & 3.00 & 3.00 & 2.91 & 3.00 & 3.00 & 3.00 & 0.02 \\
$3$--$5\times10^{-2}$ & 2.87 & 2.96 & 2.99 & 2.99 & 2.88 & 2.97 & 2.99 & 2.99 & 0.02 \\
$5$--$6\times10^{-2}$ & $2.8\times10^{4}$ & $2.8\times10^{4}$ & $3.0\times10^{4}$ & 2.97 & 22.5 & 22.6 & 23.5 & 2.96 & 0.02 \\
$6$--$7\times10^{-2}$ & $8.4\times10^{5}$ & $6.2\times10^{5}$ & $6.6\times10^{5}$ & 3.02 & 369 & 277 & 288 & 3.02 & 0.02 \\
$7$--$8\times10^{-2}$ & 2.23 & 2.30 & 2.99 & 2.99 & 2.64 & 2.72 & 2.99 & 2.99 & 0.02 \\
$8$--$9\times10^{-2}$ & 11.3 & 11.6 & 26.5 & 2.98 & 3.01 & 3.10 & 3.85 & 2.97 & 0.02 \\
$9\times10^{-2}$--$10^{-1}$ & 105 & 106 & 254 & 3.02 & 8.99 & 9.11 & 14.5 & 3.01 & 0.02 \\
\midrule
\SI{30}{mHz} zero, $\pm 20$ bins & 3.34 & 3.32 & 3.33 & 3.33 & 3.48 & 3.45 & 3.45 & 3.30 & 0.27 \\
\SI{60}{mHz} zero, $\pm 20$ bins & $4.2\times10^{6}$ & $4.1\times10^{6}$ & $4.3\times10^{6}$ & 2.66 & $2.4\times10^{3}$ & $2.3\times10^{3}$ & $2.4\times10^{3}$ & 2.76 & 0.27 \\
\SI{90}{mHz} zero, $\pm 20$ bins & $7.1\times10^{3}$ & $7.1\times10^{3}$ & $1.7\times10^{4}$ & 3.21 & 477 & 480 & 675 & 3.23 & 0.27 \\
\bottomrule
\end{tabular}
\end{table*}

\section{The real and imaginary parts of the cross spectra}\label{app:real_imag_cross_spectra}

\begin{figure*}[t]
    \centering
    \includegraphics[width=0.86\textwidth]{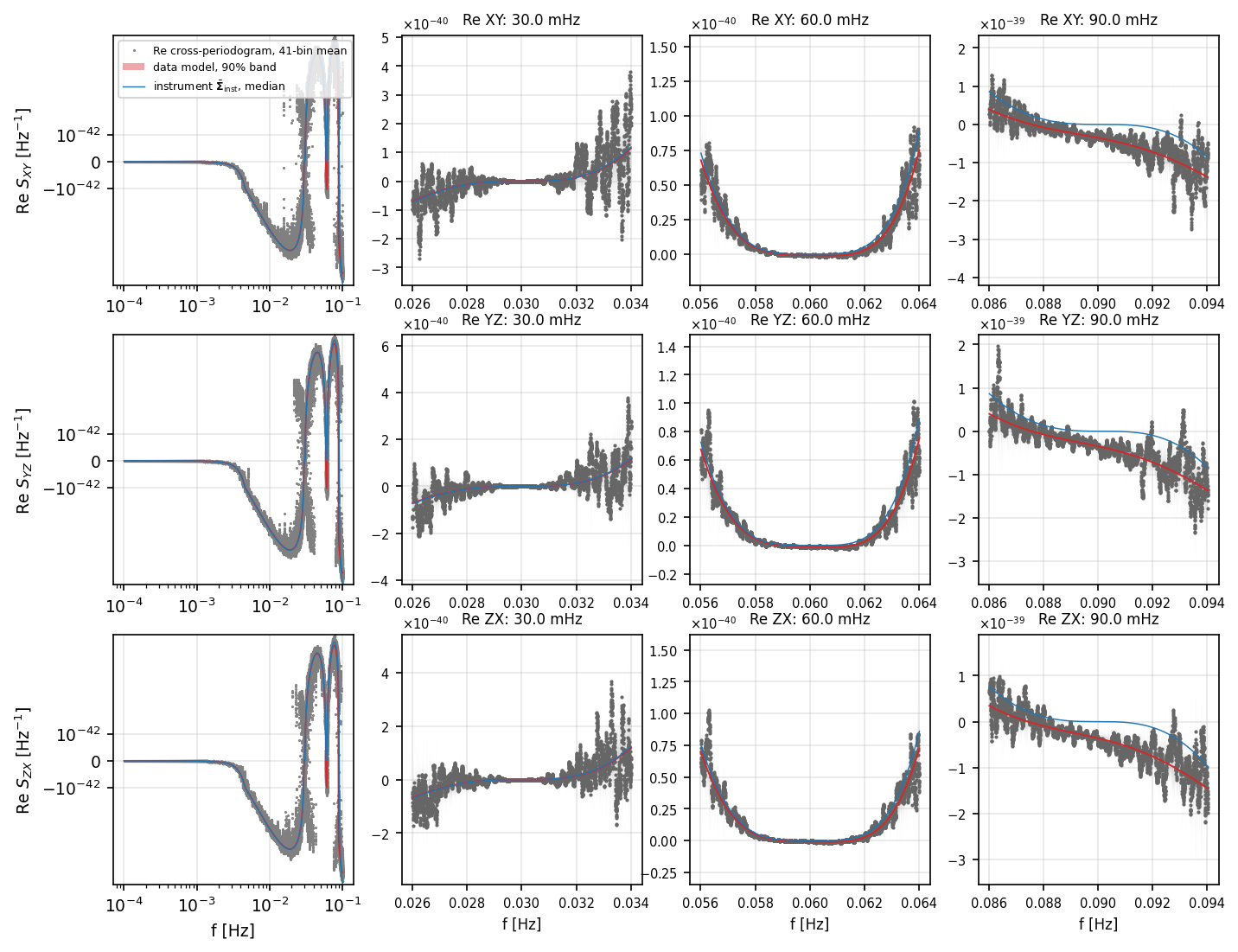}
    \caption{The posterior cross-spectra of the first segment. \emph{Left}: the real parts of the $XY$, $YZ$ and $ZX$ cross-spectra, the cross-periodogram averaged over $41$ bins (grey points) against the posterior median and $90\%$ band of the data model (red) and the median of the instrument covariance (blue). \emph{Right}: the same within $\pm\SI{4}{mHz}$ of each zero, on a linear scale, with the $\pm 1\sigma$ scatter of the average drawn about the median. The magnitude of a cross-periodogram is not an unbiased estimate of $|C_{ij}|$ where the coherence is low, which is why the real part is compared. Produced by \srclink{analyse.py}.}
    \label{fig:mcmc_bands_csd}
\end{figure*}

In Fig.~\ref{fig:mcmc_bands_csd} we show the real parts of the cross-spectra. Above \SI{10}{mHz} the auto- and cross-spectra are determined to a few per cent, and the bands stay that narrow through the bottoms of the zeros. The depth of a zero is set by the leaked and aliased power of its shoulders rather than by any knot. The knot value bands widen to tens of per cent at the low edge of the band due to the limited amount of knots and data. The cross-spectra are a check of the geometry of LISA rather than a fit of the noise process. The imaginary components contain no free parameters beyond the seventeen knots that the auto-spectra already fix, and the model follows the cross-periodogram across the band and through the zeros. Only the real parts are shown, even though the model retains the imaginary parts. For equal arms the cross-spectra are real, and the imaginary parts that the unequal arms induce are a small fraction of the real ones. An analysis that chose to neglect them, as a frequency-domain spline fit to the auto- and cross-spectra might, would lose a little information and essentially zero accuracy.

To see this, write the true covariance
\begin{equation}\bSigma = \boldsymbol{A} + i\boldsymbol{B}\end{equation}
with $\boldsymbol{A}$ symmetric and $\boldsymbol{B}$ antisymmetric. This is easily verified since $\bSigma = \bSigma^{\dagger}$ and equating real and imaginary parts. Now, let $\bSigma^{\prime}(\blambda) = {\rm Re}\,\bSigma(\blambda)$ denote the model that only contains the real parts, such that $\bSigma^{\prime} = \bA$ at the true knot values. In the notation of Ref.~\cite{burke2026mindsecondgapimpact}, with $\bQ = \bSigma$ the true covariance and $\bSigma^{\prime}_{\beta} = \partial_{\beta}\bSigma^{\prime}$, the leading-order noise-induced fluctuation of the recovered knot values and its expectation, the systematic shift, are [Eqs.~(74) and (71) there]\footnote{The following expressions are time-domain quantities rather than frequency domain. This has no impact on our conclusions -- the Discrete Fourier Transform matrix is a unitary matrix so traces are equivalent in any time or frequency domain. For example, $\text{Tr}(\Sigma) = \text{Tr}(\bP^{\dagger}\tilde{\bSigma}\bP) = \text{Tr}(\tilde{\bSigma})$.}
\begin{widetext}
\begin{align}
    \Delta\hat{\lambda}^{\alpha} &= \frac{1}{2}(\boldsymbol{\Gamma}^{\prime})^{-1}_{\alpha\beta}\Big[(\bW\bn)^T(\bSigma^\prime)^+ \bSigma^\prime_{\beta}(\bSigma^\prime)^+ (\bW\bn) - \text{Tr}[(\bSigma^\prime)^+ \bSigma^\prime_{\beta}]\Big]\,, \qquad
    \boldsymbol{\Gamma}^{\prime}_{\alpha\beta} = \frac{1}{2}\text{Tr}\big((\bSigma^\prime)^+ \bSigma^\prime_{\alpha}(\bSigma^\prime)^+\bSigma^\prime_{\beta}\big)\,, \\
    \langle \Delta\hat{\lambda}^\alpha \rangle &\approx \frac{1}{2}(\boldsymbol{\Gamma}')^{-1}_{\alpha\beta}\;\text{Tr}\!\left[(\boldsymbol{\Sigma}')^+\boldsymbol{\Sigma}'_\beta\left((\boldsymbol{\Sigma}')^+\bSigma - \mathbb{I}\right)\right]
    = \frac{i}{2}(\boldsymbol{\Gamma}')^{-1}_{\alpha\beta}\;\text{Tr}\big(\boldsymbol{S}_{\beta}\bB\big)\,, \qquad \boldsymbol{S}_{\beta} \equiv (\bSigma^{\prime})^{+}\bSigma^{\prime}_{\beta}(\bSigma^{\prime})^{+}\,,
\end{align}
\end{widetext}
the last equality holding at the true knot values, where $\bQ - \bSigma^{\prime} = i\bB$. Observe now that for the real antisymmetric matrix $\bB = -\bB^{\mathrm{T}}$, the cyclic property $\text{Tr}(\boldsymbol{M}\boldsymbol{N}) = \text{Tr}(\boldsymbol{N}\boldsymbol{M})$ and $\boldsymbol{S}_{\beta}$ real symmetric, we obtain
\begin{equation}
\text{Tr}(\boldsymbol{S}_{\beta}\bB) = \text{Tr}([\boldsymbol{S}_{\beta}\bB]^{\mathrm{T}}) = \text{Tr}(\bB^{\mathrm{T}}\boldsymbol{S}_{\beta}) = -\text{Tr}(\boldsymbol{S}_{\beta}\bB)\,,
\label{eq:tr_antisym}
\end{equation}
indicating that $\langle \Delta\hat{\lambda}^\alpha \rangle = 0$ for every knot, at every bin and for any geometry. The covariance of this quantity is given in Eqs.~(76)--(78) of Ref.~\cite{burke2026mindsecondgapimpact} 
\begin{widetext}
\begin{align}
    \langle\Delta\hat{\lambda}^\alpha\Delta\hat{\lambda}^\beta\rangle = (\boldsymbol{\Gamma}')^{-1}_{\alpha\rho}(\boldsymbol{\Gamma}')^{-1}_{\beta\sigma}\,\frac{1}{2}\text{Tr}\!\left[\boldsymbol{S}_{\rho}\bQ\,\boldsymbol{S}_{\sigma}\bQ\right]
    = (\boldsymbol{\Gamma}')^{-1}_{\alpha\beta} - (\boldsymbol{\Gamma}')^{-1}_{\alpha\rho}(\boldsymbol{\Gamma}')^{-1}_{\beta\sigma}\,\frac{1}{2}\text{Tr}\!\left[\boldsymbol{S}_{\rho}\bB\,\boldsymbol{S}_{\sigma}\bB\right]\,,
\end{align}
\end{widetext}
the bias-product term of the scatter vanishing by Eq.~\eqref{eq:tr_antisym} and the cross terms linear in $\bB$ cancelling by the same identity, so that the scatter differs from the quoted width only at second order in $\bB$. Notice that the first term is precisely the model Fisher matrix. Since $\bB$ is a small fraction of $\bA$ for the geometries related to LISA, this latter term is vanishingly small. 

Following Ref.~\cite{burke2026mindsecondgapimpact}, let $\Upsilon_{\alpha} = \sqrt{\langle\Delta\hat{\lambda}^\alpha\Delta\hat{\lambda}^\alpha\rangle / (\boldsymbol{\Gamma}')^{-1}_{\alpha\alpha}}$ be the scatter of knot $\alpha$ in units of the width the real-only model quotes, and $\Xi_{\alpha} = \sqrt{(\boldsymbol{\Gamma}')^{-1}_{\alpha\alpha} / \boldsymbol{\Gamma}^{-1}_{\alpha\alpha}}$ that width in units of the width of the full model. On segment 0 we have computed $\Upsilon_{\alpha}$ and the result is between $0.9999$ and $1.0002$ and $\Xi_{\alpha}$ between $1.0000$ and $1.0024$; at the most asymmetric constellation of two years of the orbits, $0.999$--$1.002$ and $1.000$--$1.030$. We have checked the information loss in \srclink{imaginary_parts.py}, where the knot uncertainties widen by at most three per cent even at the most asymmetric constellation of two years of the orbits. The points here are critical if one were to assume a more model agnostic free spline fitted to the auto and cross spectra of LISA. One could build splines for the real parts and neglect the imaginary parts, reducing the number of parameters significantly and still yielding unbiased and accurate noise inference.

\section{The sampled posterior of the knots}
\label{app:corner}

\begin{figure*}[p]
    \centering
    \includegraphics[width=0.95\textwidth]{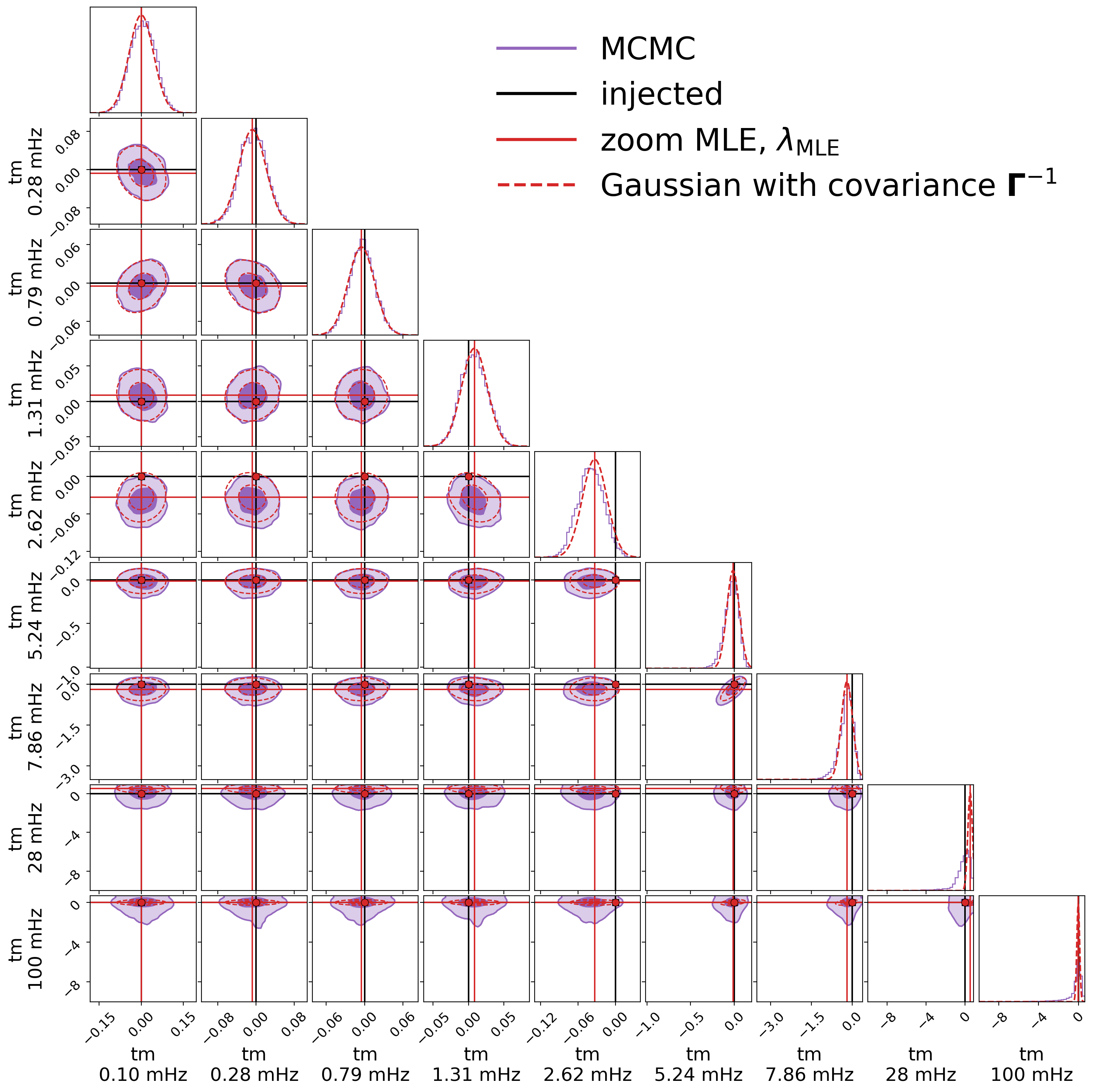}
    \caption{The nine test-mass knots, in $\log_{10}$ from the injected value, which is the black line and the red dot. The posterior is purple. The zoom maximum $\lambda_{\rm MLE}$ is the red line. The red dashes are a multivariate Gaussian centred on the MLE, with covariance given by the inverse Fisher matrix $\bGamma^{-1}$, drawn as its marginal on the diagonal and its $1\sigma$ and $2\sigma$ contours on the other panels. Where the red dashes lie on the purple, the Gaussian describes the posterior. The seven knots below \SI{10}{mHz} are centred on the injected values; the two above have a sharp upper edge and trail off towards lower values. Produced by \srclink{fig_corner.py}.}
    \label{fig:corner_tm}
\end{figure*}

\begin{figure*}[p]
    \centering
    \includegraphics[width=0.95\textwidth]{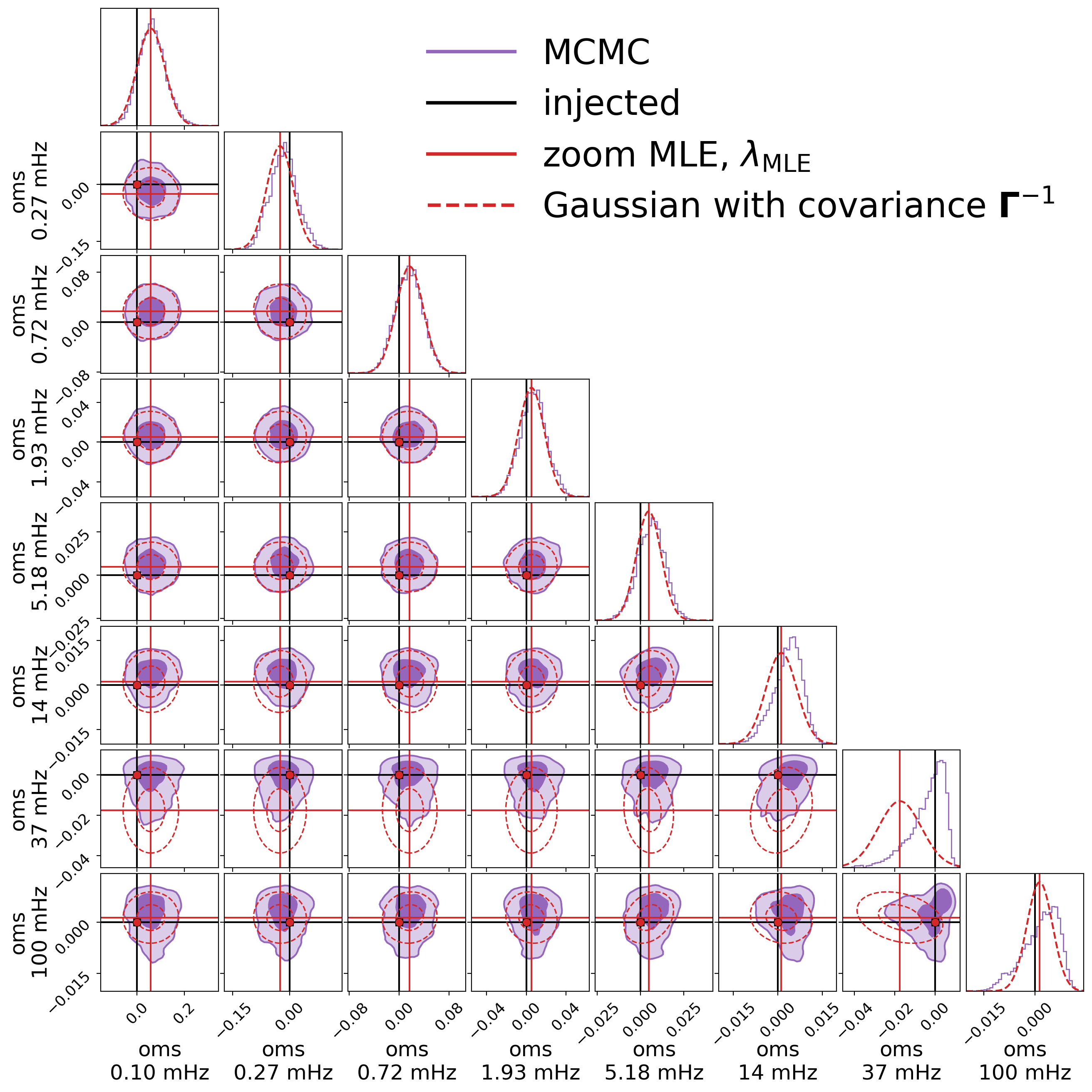}
    \caption{The eight readout knots, marked as in Fig.~\ref{fig:corner_tm}. The posterior of every knot is centred on the injected value to within $0.06$ in $\log_{10}$ at \SI{0.1}{mHz} and $0.005$ from \SI{5}{mHz} up. Produced by \srclink{fig_corner.py}.}
    \label{fig:corner_oms}
\end{figure*}

Figures~\ref{fig:corner_tm} and \ref{fig:corner_oms} show the posterior of Sec.~\ref{subsec:results_mcmc}. Every axis is the difference of the recovered knot value from the injected one in units of  $\log_{10}$. The contours enclose $39\%$ and $86\%$ of the samples, the $1\sigma$ and $2\sigma$ levels of a two-dimensional Gaussian. We overlay a Gaussian distribution centred on the MLE with width predicted by the inverse Fisher matrix computed through the Hessian of the log-likelihood function. The MCMC and MLE agree exceptionally well where the posterior samples are Gaussian. 

\clearpage

\bibliographystyle{apsrev4-2}
\bibliography{gremlins}

\end{document}